\documentclass[12pt]{article}
\usepackage{amsmath}
\usepackage{graphicx}
\usepackage[authoryear]{natbib}
\usepackage{url} 
\usepackage[utf8]{inputenc}
\usepackage{lmodern}
\usepackage[T1]{fontenc}
\usepackage{amsmath,amsthm,amsfonts,amssymb,bm,mathtools}
\usepackage{dsfont}			
\usepackage{enumitem}
\usepackage{graphicx}
\usepackage{booktabs}
\usepackage[format=plain,labelfont=it,textfont=it]{caption}
\usepackage{xcolor}
\usepackage[colorlinks,linkcolor=blue,citecolor=blue,urlcolor=blue]{hyperref}
\newcommand{\blind}{0}

\newcommand{\X}{\mathbf{X}}

\newcommand{\SIGMA}{\bm{\Sigma}}
\newcommand{\OMEGA}{\bm{\Omega}}
\newcommand{\PSI}{\bm{\Psi}}
\newcommand{\THETA}{\bm{\Theta}}
\newcommand{\GAMMA}{\bm{\Gamma}}

\newcommand{\PP}{\mathbf{P}}
\newcommand{\M}{\mathbf{M}}

\DeclareMathOperator*{\tr}{tr}
\DeclareMathOperator*{\sign}{sign}

\DeclareMathOperator*{\minimize}{minimize}

\begin{document}

\def\spacingset#1{\renewcommand{\baselinestretch}%
{#1}\small\normalsize} \spacingset{1}

\newcommand{\TODO}[1]{\textcolor{red}{TODO-MF: #1}}
\newcommand{\NOTE}[1]{\textcolor{magenta}{NOTE-MF: #1}}


\if0\blind
{%
  \title{\bf Sparse model-based clustering of three-way data via lasso-type penalties}
  \author{Andrea Cappozzo\footnote{These authors contributed equally to this work}
    \hspace{.2cm}\\
    MOX, Department of Mathematics, Politecnico di Milano\\
    Alessandro Casa$^*$
    \hspace{.2cm}\\
    Faculty of Economics and Management, Free University of Bozen-Bolzano \\
    Michael Fop\\
    School of Mathematics \& Statistics, University College Dublin}
  \maketitle
} \fi

\if1\blind
{
  \bigskip
  \bigskip
  \bigskip
  \begin{center}
    {\LARGE\bf Title}
\end{center}
  \medskip
} \fi

\bigskip
\begin{abstract}

Mixtures of matrix Gaussian distributions provide a probabilistic framework for clustering continuous matrix-variate data, which are becoming increasingly prevalent in various fields. Despite its widespread adoption and successful application, this approach suffers from over-parameterization issues, making it less suitable even for matrix-variate data of moderate size. To overcome this drawback, we introduce a sparse model-based clustering approach for three-way data. Our approach assumes that the matrix mixture parameters are sparse and have different degree of sparsity across clusters, allowing to induce parsimony in a flexible manner. Estimation of the model relies on the maximization of a penalized likelihood, with specifically tailored group and graphical lasso penalties. These penalties enable the selection of the most informative features for clustering three-way data where variables are recorded over multiple occasions and allow to capture cluster-specific association structures. The proposed methodology is tested extensively on synthetic data and its validity is demonstrated in application to time-dependent crime patterns in different US cities. 
\end{abstract}

\noindent%

\spacingset{1.5} 
\section{Introduction} \label{sec:intro}

Matrix-variate data, where a matrix is observed for each statistical unit, are becoming more common in a large number of applications and data analysis routines. This data structure is often referred to as {\em three-way} and characterized by the presence of three different layers or modes, namely the units, the variables and the occasions. These data are nowadays often occurring in applications such as multivariate time-dependent analysis \citep{anderlucci2015covariance}, the analysis of crime patterns \citep{Melnykov2019a}, basketball analytics \citep{yin:2023}, the analysis of export trade networks \citep{melnykov:2021:network},   image and brain scan data analysis \citep{gao2021regularized,liu:2022}. In spite of their potential in terms of informative content, matrix-variate data introduce several challenges which need to be dealt with in the modeling process. In fact, each of the three different layers induce specific peculiarities in terms of intricate dependency structures.

In this landscape, clustering is often of interest to reduce the aforementioned complexities by proposing parsimonious summaries of the data and highlighting their most relevant patterns. To this extent, both distance-based \citep{vichi1999one,vichi2007simultaneous} and nonparametric techniques \citep{ferraccioli2022modal} have been proposed. Nevertheless, parametric or model-based approaches are undoubtedly the ones that have received the most attention: taking steps from \citet{basford1985mixture} and building on mixtures of matrix-variate Gaussian distributions, the seminal papers by \citet{viroli:2011,viroli:2011:b} have paved the way for a new and lively stream of research. Recently, several flexible approaches have been proposed to deal with data of different nature. These approaches considered alternative distributional assumptions for skewed data \citep{chen2005matrix,melnykov2018model,gallaugher2018finite}, transformations \citep[see, among others,][]{chen2005matrix,melnykov2018model,gallaugher2018finite,tomarchio2020two,Tomarchio2022,tomarchio2022matrix} and alternative models for count data \citep{silva2023finite,subedi:2023}. 

Despite being practically useful, matrix-variate model-based clustering faces significant limitations in high-dimensional settings. These limitations are particularly pronounced in the three-way framework where the tendency to over-parameterization, inherited from the vector-valued setting \citep{bouveyron:2014}, becomes even more challenging. In fact, in the context of the matrix Gaussian distribution, two covariance matrices are employed for each component to accommodate the data structure. Consequently, when dense parameterizations are assumed for these matrices, the number of parameters to be estimated grows quadratically with both the number of rows and columns. This undermines the practical utility of the approach, even when a moderate number of variables and/or occasions are observed. 

To address these limitations, in this work we introduce a novel approach where each parameter involved in the specification of the matrix Gaussian mixture model has its own cluster-specific degree of sparsity. This greatly increases the flexibility of the model, leads to a parsimonious modeling framework, and provides more interpretable insights regarding the clustering partition. The approach relies on the maximization of a penalized likelihood which automatically enforces sparsity. More specifically, we impose a graphical lasso penalty on rows and columns precision matrices, promoting a reduction in the number of non-zero parameters while facilitating interpretation in terms of conditional dependencies, thanks to the connection with Gaussian graphical models. Additionally, we impose a group lasso penalty on the rows of the component mean matrices. In the common scenario where variables are observed over time for a set of statistical units, this penalization scheme allows to perform automatic variable selection in a three-way model-based clustering framework. As a supplementary contribution, we briefly generalize the applicability of the work by \citet{heo2021penalized}, where they consider a lasso-type entry-wise penalty for the elements of the mean matrices. 

The remainder of the paper is structured as follows. Section \ref{sec:mbcmatrix} overviews model-based clustering of matrix-variate data, with a specific focus on the issues arising in high-dimensional spaces. In Section \ref{sec:sparsemixmat}, our proposal is introduced and motivated, alongside with the description of the associated estimation and model selection methods. In Section \ref{sec:sim_study} and \ref{sec:crimeApplication}, the performance of the proposed framework  is tested on synthetic and real data, respectively. Conclusions and considerations about further improvements and future research directions end the paper in Section \ref{sec:conc}. 

\newpage

\section{Model-based matrix-variate clustering}\label{sec:mbcmatrix}

\subsection{Mixture of matrix normal distributions} 	\label{sec:GMM_primer}
Model-based clustering \citep{fraley:raftery:2002,bouveyron2019model} assumes that the data are generated by a finite mixture distribution, which describes the presence of heterogeneous sub-populations. In this context, typically maximum likelihood estimation is usually implemented by means of the EM algorithm \citep{dempster:etal:1977}, resorting to a data augmentation scheme where the latent group indicator variables are treated as missing data. Operationally, once the model is fitted, a partition is obtained by assuming a one-to-one correspondence between the groups and the mixture components, and assigning the $i$-th observation to a given cluster according to the maximum a posteriori (MAP) rule \cite[see][for a detailed treatment]{fraley:raftery:2002,bouveyron2019model}. 

When dealing with standard continuous vector-variate data, where a number of variables are measured for a set of units, it is routine to assume that the mixture components correspond to multivariate Gaussian distributions \citep{fraley:raftery:2002}. Nonetheless, nowadays it is becoming increasingly common to encounter three-way data structures, where multiple variables are measured over different occasions. This additional layer (or mode) introduces new modeling challenges that need to be taken into account when clustering samples is the final goal. Indeed, as noted by \citet{anderlucci2015covariance}, models have to ``{\em account simultaneously for three goals of the analysis, which arise from the three layers of the data structure; heterogeneous units, correlated occasions and dependent variables}''. 
Matrix Gaussian mixture models have originally been proposed by \citet{viroli:2011,viroli:2011:b} with the aim of simoultaneously accounting for these sources of complexity.

Formally, let $\X = \{\X_1, \dots, \X_n \}$, be a sample of $p\times q$ matrices, with $\X_i \in \mathbb{R}^{p \times q}$, $i=1,\ldots,n$. While in general the dimensions can be relate to any type of measurement, in the following we assume that $p$ variables are observed in $q$ different occasions, as appropriate for most applications. The natural GMM extension for model-based clustering of three-way data is given by the matrix Gaussian mixture model (MGMM), expressed as follows:
\begin{eqnarray}\label{eq:matrixnormalmixture}
f(\X_i; \bm{\Theta}) = \sum_{k=1}^K \tau_k \phi_{p\times q}(\X_i; \M_k, \SIGMA_k, \PSI_k), 
\end{eqnarray}
where $\tau_k$'s are the mixing proportions with $\tau_k > 0$, $\forall k = 1, \dots, K$ and $\sum_{k = 1}^K \tau_k = 1$; $K$ is the number of mixture components, while $\THETA$ denotes the collection of all mixture parameters. Here, $\phi_{p\times q}(\cdot; \M_k, \SIGMA_k, \PSI_k)$ denotes the density of a $p \times q$ matrix normal distribution \citep{dawid:1981}, reading as
\begin{eqnarray*}
 \phi_{p\times q}(\X_i; \M_k, \SIGMA_k, \PSI_k) &=& (2\pi)^{-\frac{pq}{2}}|\SIGMA_k|^{-\frac{q}{2}}|\PSI_k|^{-\frac{p}{2}}  \\
 && \exp\left\{-\frac{1}{2}\text{tr}(\SIGMA_k^{-1}(\X_i - \M_k)\PSI_k^{-1}(\X_i - \M_k)') \right\},
\end{eqnarray*}
where $\M_k$ is the $p\times q$ mean matrix of the $k$-th component, and $\SIGMA_k$ and $\PSI_k$ are the component rows and columns covariance matrices, with dimensions $p\times p$ and $q \times q$, respectively. Coherently with the two-way scenario, the model in (\ref{eq:matrixnormalmixture}) can be estimated by means of the EM-algorithm, see for example \citet{viroli:2011,Glanz2018,gao2021regularized}. Alternatively, the model can also be formulated and estimated under a Bayesian framework, as for example \citet{viroli:2011,yin:2023}.

An alternative specification of the matrix-variate Gaussian distribution may be given, since the following relation holds
\begin{equation} \label{eq:GMM_MGMM}
\X \sim m\mathcal{N}_{p\times q}(\M, \SIGMA, \PSI) \iff \text{vec}(\X) \sim \mathcal{N}_{pq}(\text{vec}(\M),  \PSI \otimes \SIGMA), 
\end{equation}
where $\text{vec}(\cdot)$ and $\otimes$ denote respectively the vectorization operator and the Kronecker product; $m\mathcal{N}_{p\times q}$ denotes a matrix Normal distribution of dimensions $p$ and $q$. From this relation, the matrix-variate Gaussian can be regarded as a direct generalization of the normal distribution to the three-way matrix framework. For more details about the matrix Gaussian distribution, its properties, and its connection to the multivariate normal distribution, readers can refer to \citet{gupta2018matrix}. The presence of the Kronecker product in \eqref{eq:GMM_MGMM} highlights an identifiability issue, since $\PSI \otimes \SIGMA = c\PSI \otimes c^{-1}\SIGMA$ for any $c \in \mathbb{R}^+$. Enforcing constraints on the trace or on the determinant of one of the two matrices is regarded as a viable solution to solve the problem \citep[see e.g.,][]{viroli2012matrix,melnykov2018model, Glanz2018}; the latter approach will be considered in the rest of the paper.


\subsection{Issues in matrix mixture models for high-dimensional data}\label{sec:highDimMatrixClust}
Finite mixture models are routinely used for probabilistic cluster analysis. Nonetheless, both in the two-way and the three-way framework, they present a cumbersome issue which is related to their tendency to be over-parameterized even with a moderate number of variables. When dealing with vector-variate data, the cardinality of the parameter space $|\THETA|$ scales quadratically with the number of variables; this problem is even more exacerbated in the matrix-variate scenario, where $|\THETA|$ scales quadratically with both dimensions $p$ and $q$ of the component row and column covariance matrices. In order to deal with this challenge, different approaches have been proposed in the two-way setting \citep[see e.g.,][for exhaustive reviews of the topic]{bouveyron:2014,fop2018variable}. which can be grouped into three distinct types: constrained modeling, variable selection, and sparse modeling; a brief overview is provided in \cite{casa2022group}. 

In line with this classification, recent efforts have been devoted to addressing the issue of over-parameterization within the framework of matrix mixture modeling. Specifically, some of the existing approaches either adopt parsimonious parametrizations, or implement variable selection to discard irrelevant variables and reduce the number of parameters. In \citet{sarkar:2020}, the authors extend the family of covariance eigendecomposition models considered for vector-valued data \citep{Banfield1993,Celeux1995} to the matrix-variate scenario. They introduced a collection of 98 constrained models and further enhanced parsimony by proposing an additive formulation for the mean matrices, resulting in a family of 196 matrix mixture models. On the other hand, \citet{wang:2020} propose a variable selection approach where the work by \citet{maugis2009variable} is extended to the matrix-variate framework. A stepwise variable selection procedure is proposed, which alternates variable inclusion and exclusion steps, where the resulting models are compared by means of the Bayesian Information Criterion \citep[BIC,][]{schwarz1978estimating}. 
These two approaches present some relevant drawbacks: they can be computationally intensive, since involve fitting and comparing of a large number of models, and they 
implement a rigid way to induce parsimony, not allowing the association structures among the variables and the structure of the mean matrices to vary from one cluster to the other.  

For the above reasons, in this work we take a different perspective, based on the formulation of a sparse matrix mixture model, by extending the framework of sparse and penalized mixture models \citep[][among others]{fop2018variable,fop:2019,casa2022group} to matrix-variate data. Building primarily upon the literature on sparse matrix graphical models \citep[][for example]{leng:2012,chen:2019} and sparse model-based clustering \citep{zhou:2009}, sparse approaches for matrix-variate data clustering have been recently introduced and are gaining increasing attention. In application to brain imaging data, \citet{gao2021regularized} develop a penalized Gaussian matrix mixture model, where penalty functions on the entries of the component mean matrices are introduced to shrink the mean parameters. The method is shown to recover the low rank mean signal, however, it does not allow a flexible modeling of the association structure across the clusters. On a similar vein, \citet{liu:2022} presents a multi-step approach for clustering and sparse correlation estimation in application to functional magnetic resonance imaging data. Here, in contrast to \citet{gao2021regularized} and motivated by the application, the authors propose an optimization framework that focuses on recovering the different association structures across the clusters, but covariance parameters rather than the means are employed to cluster the units, which could be a limitation if clusters are well separated in terms of mean signals. Additionally, the authors remark that the method suffer from the need to pre-specify the number of clusters beforehand and the lack of a principled method for selecting this number. In \citet{heo2021penalized}, the authors describe a penalized matrix normal mixture model for clustering that employs penalty functions on both means and covariance matrix parameters to induce sparse estimation. However, this approach relies on implicit restrictive independence assumptions during estimation, posing potential problems. Moreover, the specific formulation of the penalty functions on the mean parameters does not allow for an effective variable selection in the context of three-way data where variables are measured over multiple occasions. 

In what follows, we propose a sparse matrix Gaussian mixture model where we overcome the drawbacks of the aforementioned frameworks for three-way data clustering. Our proposed approach offers several advantages: it allows clusters to be characterized by different association structures, it accommodates estimation of sparse component matrix means and inverse covariance matrices, it uses a principled criterion for model selection, it leverages a computationally efficient framework for estimation based on lasso-type penalties, it allows mean parameters to have different sparse patterns across clusters, and it implements variable selection in a matrix-variate context where the variables are observed over multiple time occasions. The proposed method is based on a maximum penalized likelihood framework, presented in the next section.

\section{Sparse matrix mixture models} \label{sec:sparsemixmat}
\subsection{Model specification} \label{sec:grLasso_modSpec}
A sparsity-inducing procedure relying on a penalized likelihood estimation approach is hereafter proposed. Following from the model in \eqref{eq:matrixnormalmixture}, we aim to maximize the general penalized log-likelihood below: 
\begin{eqnarray}\label{eq:pen_likelihood}
\ell_P(\THETA; \X) = \sum_{i=1}^n \log\left\{\sum_{k=1}^K \tau_k \phi_{p\times q}(\X_i; \M_k, \OMEGA_k, \GAMMA_k) \right\} - p_{\bm\lambda}(\M_k,\OMEGA_k,\GAMMA_k)\, ,
\end{eqnarray}
where the first addend represents the standard MGMM log-likelihood and $p_{\bm\lambda}(\M_k,\OMEGA_k,\GAMMA_k)$ is a penalty term depending on a set of shrinkage factors generally denoted with ${\bm\lambda}$,
while $\OMEGA_k = \SIGMA_k^{-1}$ and $\GAMMA_k = \PSI_k^{-1}$, for $k=1,\ldots,K$ are the rows and column precision matrices, respectively. The collection of parameters is $\THETA=\left\{\tau_k ,\M_k, \OMEGA_k, \GAMMA_k\right\}_{k=1}^K$. The choice to parameterize the MGMM density in terms of precision matrices is motivated by their relation to Gaussian graphical models and their interpretation in terms of conditional dependencies \citep{whittaker:1990,leng:2012}. However, other options could be considered, and a discussion is reported in Section \ref{sec:conc}.

Different routes can be taken when specifying the penalty $p_{\bm\lambda}(\M_k,\OMEGA_k,\GAMMA_k)$ to obtain sparse estimates of the mixture component matrix parameters; readers may refer to the recent book by \citet{hastie2019statistical} for a detailed discussion. In this work, we consider the following penalty term 
\begin{eqnarray}\label{eq:penGroupLasso}
p_{\bm\lambda}(\M_k,\OMEGA_k,\GAMMA_k) = \sum_{k = 1}^K \lambda_1 \sum_{r = 1}^p \Vert \mathbf{m}_{r\cdot,k} \Vert_2 + \sum_{k = 1}^K \lambda_2 \Vert \PP_2*\OMEGA_k \Vert_1 + \sum_{k = 1}^K \lambda_3 \Vert \PP_3*\GAMMA_k \Vert_1,
\end{eqnarray}
where $\mathbf{m}_{r\cdot,k}$ is the $r$-th row of matrix $\M_k$, while $\Vert \cdot \Vert_1$ and $\Vert \cdot \Vert_2$ are the $L_1$ and the $L_2$-norm respectively, with $\Vert A \Vert_1 = \sum_{jh}|A_{jh}|$. Moreover, $\bm\lambda = (\lambda_1, \lambda_2, \lambda_3)$ is a vector of positive shrinkage hyper-parameters controlling the strength of the penalization. Lastly, $\PP_2$ and $\PP_3$ are matrices with non-negative entries, and $*$ denotes the element-wise product. 

The first term in (\ref{eq:penGroupLasso}) corresponds to a group lasso penalty \citep{yuan2006model}, imposed on the rows of the $K$ mean matrices $\M_k$. Group lasso aims to simultaneously shrink to zero a set of grouped parameters, and it has been mainly used in a regression framework, where some covariates might be structurally connected \citep[see Ch.4.3 in][and references therein]{hastie2019statistical}. In this work, we generalize this penalty to an unsupervised setting with matrix-variate mean parameters. Here, we consider the parameters as being grouped according to the rows of $\M_k$. Therefore, for a given $k$, either the whole row $\mathbf{m}_{r\cdot,k} = (m_{r1,k}, \dots, m_{rq,k})$ is estimated to be zero, or else its elements are shrunk towards zero (but not resulting equal to zero) by an amount depending on $\lambda_1$. 
This penalization scheme is adopted to perform variable selection in model-based clustering of three-way data in the common scenario when $p$ variables are observed over $q$ time instants or occasions. 
Indeed, when $\mathbf{m}_{r\cdot,k} = \mathbf{0}$ for all $k$, the $r$-th row of $\M_k$ is constant across all occasions and clusters. Therefore, it is not useful for discriminating the mean levels of the clusters. Even when $\mathbf{m}_{r\cdot,k} = \mathbf{m}_{r\cdot,h} = \mathbf{0}$ for some components $k$ and $h$, the $r$-th variable does not contain discriminative information to separate them, resulting in overlap along that dimension. 
Note that the proposed approach can be seen as the adaptation of the support union recovery methodology \citep{Obozinski2009, Obozinski2011} to the matrix-variate model-based clustering context. 

With the second and the third term in \eqref{eq:penGroupLasso}, we impose a graphical lasso penalty \citep[see][]{banerjee2008model,friedman:etal:2008,witten2011new} on the group-specific precision matrices. This represents an extension of the work by \citet{leng:2012} to the framework of mixture models. By shrinking to zero some parameters, the penalty terms allow to alleviate the problems outlined in Section \ref{sec:highDimMatrixClust} when dealing with high-dimensional data, providing a parsimonious and flexible model for the association structure between row and column variables across clusters. The resulting sparse representation of $\OMEGA_k$ and $\GAMMA_k$, for $k = 1,\dots, K$, provides a convenient interpretation of the dependencies among rows and columns of the observed matrices. In fact, zero entries in the precision matrices imply that the corresponding variables are conditionally independent given the others, following the principles of Gaussian graphical models \citep{whittaker:1990}. The matrices $\PP_2$ and $\PP_3$ in the graphical lasso penalty term introduce an higher degree of flexibility, since particular specifications allow to introduce prior beliefs regarding the dependencies between the variables. Indications on how to choose these matrices can be found in \citet{bien:tibshirani:2011}. Here the authors suggest to use all-ones matrices, ensuring homogeneous and uninformed penalization for all the precision terms. To prevent shrinkage of the diagonal entries, zeros can be placed on the main diagonal. Alternatively, $\PP_2$ and $\PP_3$ can be defined as adjacency matrices with user-defined patterns, thus allowing the a priori specification of the expected conditional dependence structures. More recently, \cite{casa2022group} introduced a data-driven method for specifying these matrices, which promotes cluster separation within the context of sparse model-based clustering and does not require initial knowledge of the association structure between the variables. In what follows we employ all-one matrices with zero diagonal entries for both $\PP_2$ and $\PP_3$, as this aspect is not the primary focus of the present paper. 

The above-mentioned methodology is based on the assumption that all the parameter matrices in (\ref{eq:pen_likelihood}), namely $\left\{ \M_k, \OMEGA_k,\GAMMA_k\right\}_{k=1}^K $, have different component-specific levels of sparsity. This leads to a realistic and flexible modeling framework, where a variable may be relevant only for a subset of clusters, and where the conditional dependence patterns are allowed to vary across groups. Our proposal represents a natural extension to the three-way data scenario of the approach outlined by \citet{zhou:2009}. Coherently with their work, the penalty on $\M_k$ aims to perform variable selection. On the other hand, the penalizations on $\OMEGA_k$ and $\GAMMA_k$ are needed in high-dimensional settings, to obtain sparse representations of the matrix mixture precision matrices and to reduce the number of free parameters to be estimated. 

\subsection{Model estimation} \label{sec:estimation}
For a fixed number of components $K$ and penalty vector $\pmb{\lambda} = (\lambda_1,\lambda_2,\lambda_3)$comprehensively, the parameters are estimated by maximizing (\ref{eq:pen_likelihood}) with respect to $\THETA$. The maximization is carried out by means of a tailored EM algorithm for maximum penalized likelihood estimation \citep{green:1990,McLachlan2008}, where the maximization step (M-step) is comprised of three partial optimization cycles. Let us firstly define the \textit{penalized complete-data log-likelihood} associated with (\ref{eq:pen_likelihood}) as
\begin{multline} \label{eq:complete_pen_ll}
\ell_C\left( \THETA; \X \right)=\sum_{i=1}^n \sum_{k=1}^K z_{ik} \left[ \log{\tau_k}-\frac{pq}{2}\log{2\pi}+\frac{q}{2}\log{|\OMEGA_k|}+\frac{p}{2}\log{|\GAMMA_k|} +\right.\\
 \left.-\frac{1}{2} \tr \left\{ \OMEGA_k \left(\X_i-\M_k \right) \GAMMA_k \left(\X_i- \M_k \right)^{'} \right\} \right]
 - p_{\pmb{\lambda}}(\M_k, \OMEGA_k, \GAMMA_k),
\end{multline}
where $z_{ik}$ is the realization of $\mathbf{Z}_{ik}$, the latent group membership indicator variable, with $z_{ik} = 1$ if matrix $\X_i$ belongs to the $k$-th component, and $0$ otherwise. The posterior probability of $\mathbf{Z}_{ik}$ is updated at each expectation step (E-step), allowing to obtain the conditional expectation of \eqref{eq:complete_pen_ll}, usually called $Q$-function, which defines the objective function to be maximized in the M-step. 
The devised algorithm is described in detail in the next subsections.

\subsubsection{Initialization strategy} \label{sec:initialization}

Initialization plays a crucial role when resorting to EM-type algorithms to perform model estimation. In fact, whenever the likelihood surface has multiple modes, the convergence to the global maximum is not guaranteed and poorly chosen initial values may lead to sub-optimal solutions \citep{McLachlan2008}. 
Thanks to the correspondence between GMM and MGMM in Equation \eqref{eq:GMM_MGMM}, initialization strategies developed for vector-variate data samples can be directly employed in the matrix-variate framework. In this regard, after the data have been vectorized, we resort to model-based agglomerative hierarchical clustering \citep{Scrucca2015}. This initialization strategy, already employed in the popular \texttt{mclust} software \citep{Scrucca2016}, 
has been proven effective in partitioning the data into $K$ initial groups. 

Once the starting partition is obtained, the first iteration of the M-step requires also initialization of the matrices $\OMEGA_k$ and $\GAMMA_k$, $k=1,\ldots, K$. For the purpose, identity matrices of dimensions respectively equal to $p \times p$ and $q \times q$ are employed as initial values.

\subsubsection{E-step}
At iteration $t$, the estimated a posteriori probabilities $\hat{z}_{i k}^{(t)}=\widehat{\operatorname{Pr}}\left(\mathbf{Z}_{ik}=1 \mid \X_{i} \right)$ are updated as follows:
$$
\hat{z}_{i k}^{(t)}=\frac{\hat{\tau}_{k}^{(t-1)} \phi_{p\times q}\left(\X_i; \hat{\M}_k^{(t-1)}, \hat{\OMEGA}_k^{(t-1)}, \hat{\GAMMA}_k^{(t-1)}\right)}{\sum_{v=1}^K \hat{\tau}_{v}^{(t-1)} \phi_{p\times q}\left(\X_i; \hat{\M}_v^{(t-1)}, \hat{\OMEGA}_v^{(t-1)}, \hat{\GAMMA}_v^{(t-1)}\right)}, \quad i=1,\ldots,n,
$$
where with the superscript $(t-1)$ we denote parameter estimates obtained in the previous EM iteration.

\subsubsection{M-step} \label{sec:mstep}
The M-step requires the maximization of the (penalized) $Q$-function, defined as
\begin{multline} \label{eq:q_func}
Q\left(\boldsymbol{\tau},\left\{\M_k, \OMEGA_k, \GAMMA_k\right\}_{k=1}^K\right)=\sum_{i=1}^n \sum_{k=1}^K \hat{z}_{i k}^{(t)}\left[ \log{\tau_k}+\frac{q}{2}\log{|\OMEGA_k|}+\frac{p}{2}\log{|\GAMMA_k|} +\right.\\\left.-\frac{1}{2} \tr \left\{ \OMEGA_k\left(\X_i-\M_k\right)\GAMMA_k\left(\X_i-\M_k\right)^{'}\right\} \right] + \\
-\sum_{k=1}^K \lambda_1 \sum_{r = 1}^p \Vert \mathbf{m}_{r\cdot,k} \Vert_2 - \sum_{k=1}^K \lambda_2 \Vert \PP_2*\OMEGA_k\Vert_1 - \sum_{k=1}^K \lambda_3\Vert \PP_3*\GAMMA_k\Vert_1.
\end{multline}
The direct maximization of $Q(\cdot)$ with respect to all parameters at once is an unfeasible task, so a partial optimization strategy is required. The closed-form expression for the mixing proportions ${\bm \tau}$ is readily available:
\begin{equation*}
\hat{\tau}_{k}^{(t)}=\frac{\hat{n}_{k}^{(t)}}{n}, \quad \hat{n}_{k}^{(t)}=\sum_{i=1}^{n} \hat{z}_{i k}^{(t)}, \quad k=1,\ldots,K.
\end{equation*}
Custom procedures are devised for obtaining updates for $\M_k$, $\OMEGA_k$, and $\GAMMA_k$, $k=1,\ldots,K$.

\subsubsection*{\normalfont\em Sparse estimation of the mean matrices $\M_k$}

When maximization of \eqref{eq:q_func} is performed with respect to $\M_k$, given current estimates of the precision matrices $\hat{\OMEGA}^{(t-1)}_k$ and $\hat{\GAMMA}^{(t-1)}_k$, the $Q$-function simplifies as follows
\begin{align} \label{eq:obj_mu}
\begin{split}
Q_{M}(\M_k)&=\sum_{i=1}^n \hat{z}_{i k}^{(t)} \left[ \tr \left\{ \hat{\OMEGA}^{(t-1)}_k\X_i\hat{\GAMMA}^{(t-1)}_k\M_k^{'}\right\}-\frac{1}{2} \tr \left\{ \hat{\OMEGA}^{(t-1)}_k\M_k\hat{\GAMMA}^{(t-1)}_k\M_k^{'}\right\} \right]- \lambda_1 \sum_{r = 1}^p \Vert \mathbf{m}_{r\cdot,k} \Vert_2\\
&=\tr \left\{ \hat{\OMEGA}^{(t-1)}_k \boldsymbol{S}_M \hat{\GAMMA}^{(t-1)}_k\M_k^{'}\right\}-\frac{\hat{n}_{k}^{(t)}}{2} \tr \left\{ \hat{\OMEGA}^{(t-1)}_k \M_k\hat{\GAMMA}^{(t-1)}_k\M_k^{'}\right\}- \lambda_1 \sum_{r = 1}^p \Vert \mathbf{m}_{r\cdot,k} \Vert_2,
\end{split}
\end{align}
where $\boldsymbol{S}_M$ is the sum of the matrix-variate observations weighted by $\hat{z}_{i k}^{(t)}$:
$$\boldsymbol{S}_M=\sum_{i=1}^n \hat{z}_{i k}^{(t)} \X_i.$$
The optimization of \eqref{eq:obj_mu} with respect to $\mathbf{M}_k$ is solved via a proximal gradient descent algorithm \citep{Parikh2014}. Briefly, proximal gradient methods address a general class of convex problems where the objective function may be decomposed into two terms: the first, generally denoted with $f(\cdot)$, is convex and differentiable, while the other, $g(\cdot)$, may not be everywhere differentiable. On that account, proximal gradient methods, also known as forward backward splitting procedures, can be seen as an extension of gradient descent for optimization problems whose gradient is not available for the entire objective function. In recent years, such approaches gained increasing popularity in the field of statistics and machine learning, as they provide reliable and numerically efficient solutions to regularized models with non-differentiable penalties \citep{Mosci2010, Klosa2020}. In our case, the maximization of \eqref{eq:obj_mu} can be recast as follows:
$$
\minimize_{\M_k} f(\M_k)+g(\M_k),
$$
where 
$$f(\M_k)=\frac{\hat{n}_{k}^{(t)}}{2} \tr \left\{ \hat{\OMEGA}^{(t-1)}_k \M_k\hat{\GAMMA}^{(t-1)}_k\M_k^{'}\right\}-\tr \left\{ \hat{\OMEGA}^{(t-1)}_k \boldsymbol{S}_M \hat{\GAMMA}^{(t-1)}_k\M_k^{'}\right\} \quad \text{and} \quad g(\M_k)=\lambda_1 \sum_{r = 1}^p \Vert \mathbf{m}_{r\cdot,k} \Vert_2.$$
Define $\nabla \mathbf{m}_{l\cdot,k}$ to be the $l$-th row, $l=1,\ldots,p$, of
\begin{equation} \label{eq:deriv_M_k}
\frac{\partial f(\M_k)}{\partial \M_k}=\hat{n}_{k}^{(t)} \hat{\OMEGA}^{(t-1)}_k \M_k \hat{\GAMMA}^{(t-1)}_k -  \hat{\OMEGA}^{(t-1)}_k \boldsymbol{S}_M \hat{\GAMMA}^{(t-1)}_k,
\end{equation}
where \eqref{eq:deriv_M_k} is the $p \times q$ matrix of first-order partial derivatives of $f(\cdot)$ with respect to $\M_k$. A proximal gradient update for the $l$-th row of matrix $\M_k$ is constructed as follows:
\begin{subequations} \label{eq:prox_grad}
\begin{align}
&\boldsymbol{b}=\mathbf{m}_{l\cdot,k}-\nu \nabla \mathbf{m}_{l\cdot,k}, \label{eq:prox_grad1}\\ 
&\hat{\mathbf{m}}_{l\cdot,k}=\text{prox}_{\nu\lambda_1}(\boldsymbol{b}), \label{eq:prox_grad2}
\end{align}
\end{subequations}
where $\nu$ is a step-size parameter and $\text{prox}_{\nu\lambda_1}(\cdot)$ is the proximity operator of the considered group lasso penalty, namely the row-wise soft thresholding operator:
\begin{equation}
\text{prox}_{\nu\lambda_1}\left(\boldsymbol{b}\right)= \begin{cases}\boldsymbol{b}\left(1-\frac{\lambda_1 \nu}{\left\|\boldsymbol{b}\right\|_2}\right)  & \text{if } \left\|\boldsymbol{b}\right\|_2>\lambda_1 \nu, \\ \boldsymbol{0} & \text{if } \left\|\boldsymbol{b}\right\|_2 \leq \lambda_1 \nu.\end{cases}
\end{equation}
Iterating equations \eqref{eq:prox_grad1} and \eqref{eq:prox_grad2} until convergence sequentially along the $p$ rows retrieves $\hat{\M}_k^{(t)}$, the estimate of the mean matrix mixture parameters for the $t$-th iteration of the EM algorithm, as the proximal gradient solution to the maximization problem in \eqref{eq:obj_mu}. When $\lambda_1$ is sufficiently large, the rows of $\hat{\M}_k^{(t)}$ are set to zero as a result of the proximity operator. 
Operationally, the weighted sample mean matrix
$\sum_{i=1}^n \hat{z}_{ik}^{(t)} \X_i/\hat{n}^{(t)}_k$
is employed as an initial guess for starting the proximal gradient search, while the step-size parameter $\nu$ is kept fixed at $10^{-4}$.

\subsubsection*{\normalfont\em Sparse estimation of the row-precision matrices $\OMEGA_k$}
When \eqref{eq:q_func} is maximized with respect to $\OMEGA_k$, given current estimates of the precision matrices $\hat{\GAMMA}^{(t-1)}_k$ and of the mean parameters $\hat{\M}_k^{(t)}$, the $Q$-function simplifies as follows:
\begin{equation} \label{eq:obj_omega}
Q_{\Omega}(\OMEGA_k)=\sum_{i=1}^n \hat{z}_{i k}^{(t)} \left[\frac{q}{2} \log{|\OMEGA_k|}-\frac{1}{2} \tr \left\{ \OMEGA_k\left(\X_i-\hat{\M}_k^{(t)}\right)\hat{\GAMMA}^{(t-1)}_k\left(\X_i-\hat{\M}_k^{(t)}\right)^{'}\right\} \right]- \lambda_2||\PP_2 * \OMEGA_k||_1.
\end{equation}
By rearranging terms in \eqref{eq:obj_omega}, we obtain:
\begin{equation} \label{eq:glasso_omega}
 Q_{\Omega}(\OMEGA_k)= \log{|\OMEGA_k|} -\tr\left\{ \OMEGA_k\boldsymbol{S}_{\Omega}\right\} -\frac{2}{\hat{n}_kq}\lambda_2||\PP_2 * \OMEGA_k||_1,
\end{equation}
where \[\boldsymbol{S}_{\Omega}=\sum_{i=1}^n \hat{z}_{i k}^{(t)}\frac{\left(\X_i-\hat{\M}_k^{(t)}\right)\hat{\GAMMA}^{(t-1)}_k\left(\X_i-\hat{\M}_k^{(t)}\right)^{'}}{\hat{n}^{(t)}_kq}.\]
Maximization of \eqref{eq:glasso_omega} with respect to $\OMEGA_k$ corresponds a graphical lasso problem, which is solved using the coordinate descent algorithm by \cite{friedman:etal:2008}, where in our context their penalty coefficient is equal to $\frac{2}{\hat{n}_kq}\lambda_2\PP_2$. The algorithm is implemented in the \texttt{R} \citep{RCoreTeam} package \texttt{glassoFast} \citep{glassoFast18} and returns the estimates of the row precision matrices $\hat{\OMEGA}_k^{(t)}$, for $k = 1, \ldots, K$.

\subsubsection*{\normalfont\em Sparse estimation of the column-precision matrices $\GAMMA_k$}
In the maximization of \eqref{eq:q_func} with respect to $\GAMMA_k$, given current estimates $\hat{\OMEGA}_k^{(t)}$ and $\hat{\M}_k^{(t)}$, the $Q$-function simplifies to:
\begin{equation} \label{eq:obj_gamma}
Q_{\Gamma}(\GAMMA_k)=\sum_{i=1}^n \hat{z}_{ik}^{(t)} \left[\frac{p}{2} \log{|\GAMMA_k|}-\frac{1}{2} \tr \left\{ \GAMMA_k\left(\X_i-\hat{\M}_k^{(t)}\right)^{'}\hat{\OMEGA}_k^{(t)}\left(\X_i-\hat{\M}_k^{(t)}\right)\right\} \right]- \lambda_3||\PP_3 * \GAMMA_k||_1.
\end{equation}
By rearranging terms in \eqref{eq:obj_gamma}, we obtain the following objective function:
\begin{equation} \label{eq:glasso_gamma}
Q_{\Gamma}(\GAMMA_k)=  \log{|\GAMMA_k|} -\tr\left\{ \GAMMA_k\boldsymbol{S}_{\Gamma}\right\} -\frac{2}{\hat{n}_kp}\lambda_3||\PP_3 * \GAMMA_k||_1, 
\end{equation}
where \[\boldsymbol{S}_{\Gamma}=\sum_{i=1}^n \hat{z}_{ik}^{(t)}\frac{\left(\X_i-\hat{\M}_k^{(t)}\right)^{'}\hat{\OMEGA}_k^{(t)}\left(\X_i-\hat{\M}_k^{(t)}\right)}{\hat{n}_k^{(t)}p}.\]
Maximization of \eqref{eq:glasso_gamma} with respect to $\GAMMA_k$ corresponds again to the graphical lasso, where in this case the original penalty coefficient is equal to $\frac{2}{\hat{n}_kp}\lambda_3\PP_3$. Also in this case the estimation is performed using the algorithm implemented in the package \texttt{glassoFast}, giving the estimates of the column precision matrices $\hat{\GAMMA}_k^{(t)}$, for $k = 1, \ldots, K$.

The updates based on the graphical lasso expressions \eqref{eq:glasso_omega} and \eqref{eq:glasso_gamma} are iterated sequentially within the M-step at each cycle of the EM algorithm until convergence is reached, returning sparse estimates of the precision matrices $\OMEGA_k$ and $\GAMMA_k$.  The global convergence is evaluated by monitoring the increase in the penalized log-likelihood at each full EM iteration. The algorithm is considered to have reached convergence when $\ell_P(\hat{\THETA}^{(t)}; \X)-\ell_P(\hat{\THETA}^{(t-1)}; \X)<\varepsilon$ for a given $\varepsilon>0$. In our analyses, $\varepsilon$ is set equal to $10^{-5}$.

The procedure described in this section is available within an \texttt{R} package at \\ \texttt{https://github.com/AndreaCappozzo/sparsemixmat}, where some of the routines have been implemented in \texttt{C++} to reduce the overall computing time.

\subsection{A note on related penalty specifications}\label{sec:Lasso_modSpec}
As briefly mentioned in Section \ref{sec:grLasso_modSpec}, several options can be considered when specifying the penalty term in (\ref{eq:pen_likelihood}). A viable alternative to our proposal would consist in considering a standard lasso penalty on the matrices $\M_k$'s, coherently with the penalty adopted for the precision matrices. In this case, the penalty term would read as follows 
\begin{eqnarray}\label{eq:penaltyLASSO_corea}
p_{\pmb{\lambda}}(\M_k, \OMEGA_k, \GAMMA_k) = \sum_{k = 1}^K \lambda_1 \Vert \PP_1 * \M_k \Vert_1 + \sum_{k = 1}^K \lambda_2 \Vert \PP_2*\OMEGA_k \Vert_1 + \sum_{k = 1}^K \lambda_3 \Vert \PP_3*\GAMMA_k \Vert_1,
\end{eqnarray}
where $\PP_1$ is a $p \times q$ matrix with non-negative entries, while the other quantities are defined as in the previous sections. Compared to the one one introduced in Section \ref{sec:grLasso_modSpec}, this penalty represents a less-structured way to induce sparsity in the mean matrices. In general, it does not allow to perform proper variable selection, since dimensions of the mean matrices are not jointly shrunk to zero. Nonetheless, the sparsity patterns could provide relevant insights and the method can be useful in some specific applications, as for example when no temporal dimension is present in the data. As highlighted in Section~\ref{sec:highDimMatrixClust}, \cite{gao2021regularized} consider lasso cell-wise penalization of matrix mixture mean parameters. However, the authors do not consider penalization of the component covariance matrices. As a result, the method may still require the estimation of a large number of parameters and does not provide a flexible model for the association structures between row and column variables.  
To overcome these limitations, in their recent work, \citet{heo2021penalized} derive a penalized matrix normal mixture model where sparsity is also induced on the precision matrices, by using a penalty function similar to \eqref{eq:penaltyLASSO_corea}. 
Nonetheless, in their proposed estimation procedure, and in particular in the M-step update for $\M_k$'s, the authors implicitly 
assume that both the rows and the columns component precision matrices are diagonal. This assumption can lead to inaccurate estimates, especially in those applications where complex conditional dependency patterns are present. For these reasons, in the following we derive an estimation scheme where the independence assumption is not required. Note that $\OMEGA_k$ and $\GAMMA_k$ are estimated as in Section \ref{sec:mstep}, therefore in what follows we only outline the updating formula for $\M_k$. Furthermore, the E-step and the considerations about the initialization strategy and the convergence criterion remain unchanged. 

Consider the current estimates of the precision matrices $\hat{\OMEGA}_k$ and $\hat{\GAMMA}_k$, where we omit the iteration superscript for ease of notation. When the penalty term is defined as in \eqref{eq:penaltyLASSO_corea}, in the maximization step with respect to $\M_k$, the $Q$-function can be expressed as follows
\begin{equation}\label{eq:Qfunct_lasso}
Q_M(\M_k)=\sum_{i=1}^n \hat{z}_{i k}^{(t)} \left[ \tr \left\{ \hat{\OMEGA}_k\X_i\hat{\GAMMA}_k\M_k^{'}\right\}-\frac{1}{2} \tr \left\{ \hat{\OMEGA}_k\M_k\hat{\GAMMA}_k\M_k^{'}\right\} \right]- \lambda_1 ||\PP_1 *\M_k||_1.
\end{equation}
We propose a cell-wise coordinate ascent estimation for $m_{ls,k}$, where $m_{ls,k}$ denotes the element in the $l$-th row and $s$-th column of matrix $\M_k$. Likewise, let $\hat{\omega}_{ls,k}$, $\hat{\gamma}_{ls,k}$ and $p_{ls,1}$ denote the elements in the $l$-th row and $s$-th column of matrices $\hat{\OMEGA}_k$, $\hat{\GAMMA}_k$ and $\PP_1$ respectively. Lastly, $x_{ls,i}$ is similarly defined in relation to a matrix observation $\X_i$. The following proposition characterizes the updating formula:\\

\textbf{Proposition 1:} \textit{The sufficient and necessary conditions for $\hat{m}_{ls,k}$ to be a (global) maximizer of \eqref{eq:Qfunct_lasso} (for fixed $l$, $s$ and $k$) are
\begin{equation} 
\sum_{i=1}^N \hat{z}_{ik}\sum_{r=1}^p\sum_{c=1}^q\hat{\omega}_{lr,k}x_{rc,i}\hat{\gamma}_{cs,k}-\hat{n}_k\sum_{r=1}^p\sum_{c=1}^q\hat{\omega}_{lr,k}\hat{m}_{rc,k}\hat{\gamma}_{cs,k}=\lambda_1 p_{ls,1}\sign(\hat{m}_{ls,k}), \quad \text{if } \hat{m}_{ls,k}\neq0
\end{equation}
and
\begin{multline}
\left| \sum_{i=1}^n \hat{z}_{ik}\left[ \sum_{\substack{r=1 \\ r\neq l}}^p \hat{\omega}_{lr,k} \left( \sum_{c=1}^q \left(x_{rc,i}-\hat{m}_{rc,k}\right) \hat{\gamma}_{cs,k} \right)+\right. \right.\\
\left. \left. + \hat{\omega}_{ll,k}\left( \sum_{\substack{c=1 \\ c\neq s}}^q \left(x_{lc,i}-\hat{m}_{lc,k}\right) \hat{\gamma}_{cs,k} \right) + \hat{\omega}_{ll,k}x_{ls,i}\hat{\gamma}_{ss,k}\right]\right|\leq \lambda_1 p_{ls,1}, \quad \text{if } \hat{m}_{ls,k}=0.
\end{multline}}
Thus, at the $t$-th iteration of the EM algorithm $\hat{m}^{(t)}_{lsk}=0$ if
\begin{multline} \label{eq:m_update_m_0}
\left| \sum_{i=1}^n \hat{z}_{ik}^{(t)}\left[ \sum_{\substack{r=1 \\ r\neq l}}^p \hat{\omega}^{(t-1)}_{lr,k} \left( \sum_{c=1}^q \left(x_{rc,i}-\hat{m}^{(t)}_{rc,k}\right) \hat{\gamma}^{(t-1)}_{cs,k} \right)+\right.\right.\\
+\left. \left. \hat{\omega}^{(t-1)}_{ll,k}\left( \sum_{\substack{c=1 \\ c\neq s}}^q \left(x_{lc,i}-\hat{m}^{(t)}_{lc,k}\right) \hat{\gamma}^{(t-1)}_{cs,k} \right) + \hat{\omega}^{(t-1)}_{ll,k}x_{ls,i}\hat{\gamma}^{(t-1)}_{ss,k}\right]\right|\leq \lambda_1 p_{ls,1},
\end{multline}
otherwise, $\hat{m}^{(t)}_{ls,k}$ is obtained by solving
\begin{eqnarray} \label{eq:m_update_m_gr_0}
\hat{n}_k^{(t)}\hat{\omega}^{(t-1)}_{ll,k}\hat{m}^{(t)}_{ls,k}\hat{\gamma}^{(t-1)}_{ss,k}+\lambda_1 p_{ls,1}\sign\left(\hat{m}_{ls,k}^{(t)}\right) &=& \sum_{i=1}^n \hat{z}_{ik}^{(t)}\sum_{r=1}^p\sum_{c=1}^q\hat{\omega}^{(t-1)}_{lr,k}x_{rc,i}\hat{\gamma}^{(t-1)}_{cs,k} + \nonumber \\ 
&& - \hat{n}_k^{(t)}\left(\sum_{\substack{r=1 \\ r\neq l}}^p\sum_{\substack{c=1 \\ c\neq s}}^q \hat{\omega}^{(t-1)}_{lr,k}\hat{m}^{(t)}_{rc,k}\hat{\gamma}^{(t-1)}_{cs,k}\right)
\end{eqnarray}
with respect to $\hat{m}^{(t)}_{ls,k}$.

The proof of Proposition 1 is reported in the Supplementary Material. This result corrects an inaccuracy introduced in Equation (5) of \cite{heo2021penalized} and it can be seen as the matrix-variate extension of Theorem 1 of \citet{zhou:2009}. Convergence to the global maximum is assured thanks to theoretical properties of coordinate descent algorithms \cite[see e.g.,][]{Wright2015}. 
The described procedure, for sufficiently large $\lambda_1$, forces some $\hat{m}_{lsk}^{(t)}$ to be shrunk to $0$, ultimately inducing sparsity in $\M_k$, $k=1,\ldots,K$. Notice however, as already mentioned, that such a penalty does not allow to directly perform variable selection within a matrix-variate data framework. The latter is achieved only employing a group-lasso penalization scheme, as highlighted in Section \ref{sec:grLasso_modSpec}. 

\subsection{Model selection} \label{sec:model_selection}
The model estimation strategy in Section \ref{sec:estimation} has been outlined by considering $K$ and $\pmb{\lambda} = (\lambda_1,\lambda_2,\lambda_3)$ fixed. However,  in practical applications, the number of clusters and the penalty hyperparameters are not known a priori and need to be chosen using model selection strategies. In this work, we select $K$ and $\pmb{\lambda}$ which maximize a modified version of the Bayesian Information Criterion \citep[BIC,][]{schwarz1978estimating}, already considered in \citet{pan:shen:2007} and \citet{casa2022group}. In detail, we use the following criterion:
\begin{equation}\label{eq:modBIC}
BIC = 2\log L(\hat{\THETA}) - d_0\log (n),
\end{equation}
where $d_0$ is the number of parameters not shrunk to zero and $\log L(\hat{\THETA})$ is the log-likelihood evaluated at $\hat{\THETA}$.

The adequacy of the BIC for selecting the number of mixture components has been thoroughly studied \citep[see e.g.,][]{roeder1997practical,keribin2000consistent}, and the criterion has been widely used both in the two-way and, more recently, in the three-way model-based clustering frameworks \citep{sarkar:2020,Tomarchio2022,Sharp2022}. Moreover, the formulation in \eqref{eq:modBIC} has been proven useful also to tune the intensity of the penalization both in the lasso \citep{zou2007degrees} and in the sparse precision matrix estimation contexts \citep{lian2011shrinkage}. Nonetheless, other model selection strategies may be pursued, especially in situations where exhaustive grid searches are considered too computational demanding. Possible alternatives are provided by stochastic optimization algorithms, such as genetic algorithms \citep{holland1992genetic}, or to conditional search schemes. Another interesting approach is outlined in \citet{jiang2015ms}, where the authors develop the E-MS algorithm, in which model selection is performed within each iteration of the standard EM algorithm. 


\section{Simulation study} \label{sec:sim_study}
\subsection{Experimental Setup} \label{sec:sim_setup}
\begin{figure}
\centering
    \includegraphics[width=\linewidth]{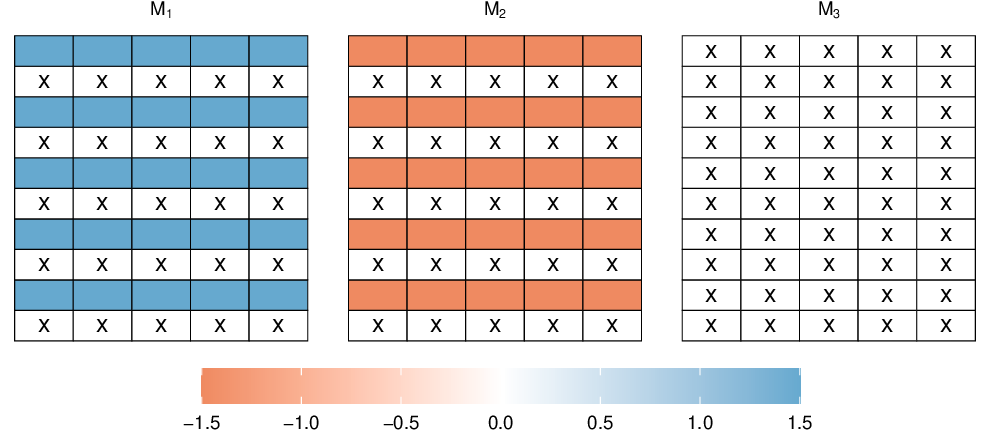}
    \caption{Heatmaps of the true $10\times 5$ mean matrices $\M_k$, $k=1,\,2,\,3$, considered in the simulated data experiment. A zero entry in the matrices is indicated with the symbol $\times$.}
  \label{fig:true_M_k_DGP}
\end{figure}

\begin{figure}
\centering
    \includegraphics[width=\linewidth]{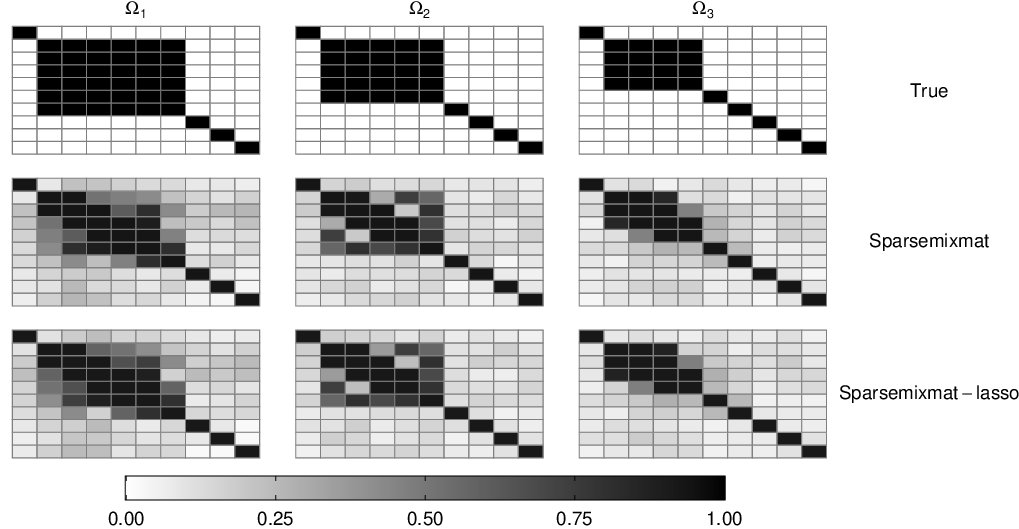}
    \caption{{\em Alternated-blocks row precision matrices} scenario. True association structures (top) and estimated association structures averaged over $100$ replications (middle and bottom) for the row precision matrices $\OMEGA_k$, $k=1,2,3$. Black squares denote a non-zero parameter between two variables.} 
  \label{fig:Omega_graph_block_cov}
\end{figure}

In this section, we assess the performance of the proposed method on synthetic data, evaluating its ability in recovering the underlying sparse patterns and the clustering structure. For each replication of the simulation experiment, we generate $n = 150$ samples from a $3$-component matrix Gaussian mixture model, in which mean matrices and both row and column precision matrices have some some degree of sparsity. The row and column precision matrices have dimensions $p \times p$ and $q \times q$, with $p$ and $q$ equal to $10$ and $5$, respectively. The $10 \times 5$ mean matrices $\M_k$, $k=1,\,2,\,3$ have a row-wise sparse structure, visually displayed in Figure \ref{fig:true_M_k_DGP}. The data generating process purposely reproduce a situation in which some of the $p$ variables measured in $q$ occasions are irrelevant for clustering. In this specific context, the second, fourth, sixth, eight and tenth row do not convey any grouping information, being identically equal to $0$ in all clusters.
We consider two distinct scenarios according to the sparsity structure enforced for the row precision matrices $\OMEGA_k$:
\begin{itemize}
\item \textit{Alternated-blocks row precision matrices:} the $10 \times 10$ row precision matrices $\OMEGA_k$, $k=1,2,3$ have a block-wise sparse structure, as visually displayed in the upper panels of Figure \ref{fig:Omega_graph_block_cov}.
\item \textit{Sparse-at-random row precision matrices:} the row precision matrices have a sparse at random Erd\H{o}s-R\'{e}nyi graph structure \citep{erdHos1960evolution} with probabilities of connection equal to $0.2$, $0.5$ and $0.8$ for $\OMEGA_1$, $\OMEGA_2$ and $\OMEGA_3$, respectively. These are visually displayed in the upper panels of Figure \ref{fig:Omega_graph_sparse_cov}.
\end{itemize}
In both scenarios, the column precision matrices $\GAMMA_k$ are generated according to a sparse at random Erd\H{o}s-R\'{e}nyi graph structure, while the mixing proportions $\tau_k$ are assumed equal to $1/K$, $K=3$. The experiment is repeated $100$ times, and for each replication the following models are fitted to the synthetic data samples:
\begin{itemize}
\item \textit{Full MGMM:} the finite mixtures of matrix normal distributions originally introduced in \citet{viroli:2011}, where full matrix parameters are estimated for each component. This model specification corresponds to a G-VVV-VV model following the nomenclature introduced in \cite{sarkar:2020}. 

\item \textit{Sparsemixmat:} the penalized MGMM method introduced in this paper, with a group-lasso penalization imposed on the rows of the mean matrices according to the penalty term in \eqref{eq:penGroupLasso}.

\item \textit{Sparsemixmat-lasso:} the penalized MGMM methodology introduced in \cite{heo2021penalized}, with a entry-wise lasso penalization on the mean matrices according to the penalty term in \eqref{eq:penaltyLASSO_corea}, and estimated following the steps outlined in Section \ref{sec:Lasso_modSpec}.
\end{itemize}

For the \textit{Sparsemixmat} and \textit{Sparsemixmat-lasso} models, a search over an equispaced grid of 
elements for each penalty term $\lambda_1$, $\lambda_2$ and $\lambda_3$ is considered, and the best model according to the BIC criterion introduced in Section \ref{sec:model_selection} is retained. 
All competing methods are initialized via model-based agglomerative hierarchical clustering as discussed in Section \ref{sec:initialization}. The methods are evaluated according to their ability in performing variable selection, recovering the true sparsity structure, and correctly retrieving the cluster allocations. The issue of matching the estimated clustering with the actual classification is addressed using the \texttt{matchClasses} function from the \texttt{e1071 R} package \citep{e1071}. Simulation results are reported in the next subsection.

\begin{figure}
\centering
    \includegraphics[width=\linewidth]{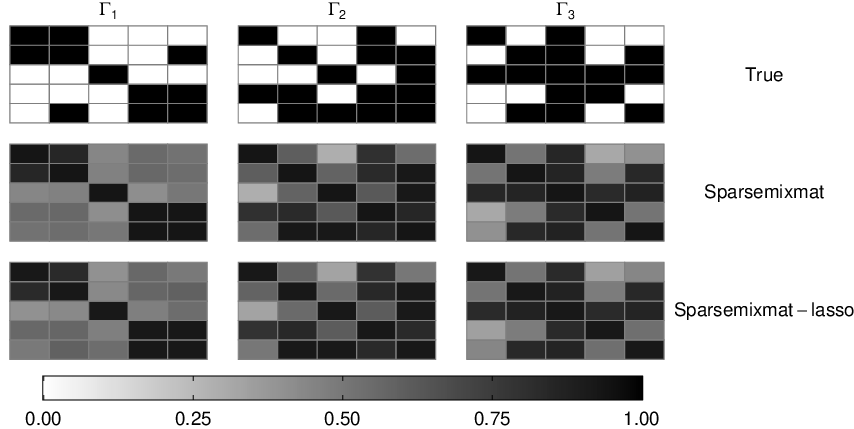}
    \caption{{\em Alternated-blocks row precision matrices} scenario. True association structures (top) and estimated association structures averaged over $100$ replications (middle and bottom) for the column precision matrices $\GAMMA_k$, $k=1,2,3$. Black squares denote a non-zero parameter between two occasions.} 
    \label{fig:Gamma_graph_block_cov}
\end{figure}

\begin{figure}
\centering
    \includegraphics[width=\linewidth]{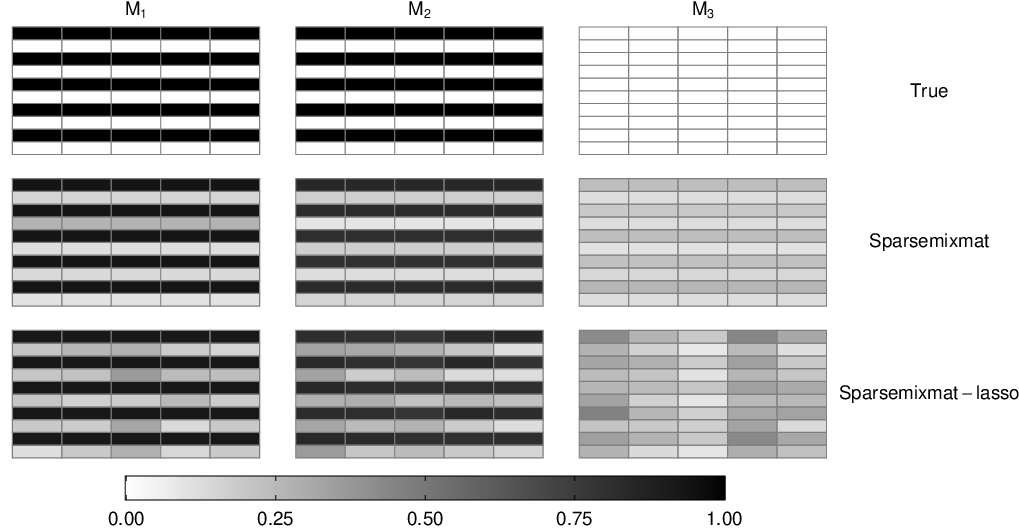}
    \caption{ {\em Alternated-blocks row precision matrices} scenario. True mean matrices (top) and estimated mean matrices averaged over $100$ replications (middle and bottom) associated to the data generating mean matrices $\M_k$, $k=1,2,3$. Black squares denote a non-zero parameter. }
  \label{fig:M_graph_block_cov}
\end{figure}

\subsection{Simulation study results}
\subsubsection{Alternated-blocks row precision matrices}
In Figure \ref{fig:Omega_graph_block_cov}, we report the heatmap plots associated to the $10 \times 10$ row precision matrices $\OMEGA_k$, $k=1,2,3$ for the \textit{alternated-blocks row precision matrices} scenario. In the top row, each heatmap represents the
association structure corresponding to a component row precision matrix, where each black square denotes the presence of a non zero parameter, and hence an association between a pair of variables. The second and third rows are the heatmaps of the proportion of times a non-zero precision parameter has been estimated between a pair of variables. As it emerges from the graphs, we note that both \textit{Sparsemixmat} and \textit{Sparsemixmat-lasso} satisfactorily recover the true underlying sparsity structure. A moderate penalty on the row-precision matrices allows for the shrinkage to zero of some of the elements of $\OMEGA_k$, which allow the correct identification of the conditional association structures among the variables in the clusters. Figure \ref{fig:Gamma_graph_block_cov} reports similar heatmaps related to the $5 \times 5$ column precision matrices. Also for this dimension of the matrix data, the association structure is correctly identified by both methods. 

Different results are observed when examining the estimates of the cluster mean matrices $\M_k$, reported in Figure \ref{fig:M_graph_block_cov}. In the figure, the heatmaps report the non-zero entries of the data generating mean matrices and the the proportion of zero entries for the estimated ones, averaged over $100$ replications. 
The row-wise shrinkage of {\em Sparsemixmat}, enforced by the group-lasso penalty, favors a better recovery of the mean matrices structure compared to the entry-wise lasso shrinkage of {\em Sparsemixmat-lasso}. This conclusion is further supported by the metrics displayed in Table \ref{tab:sim_blocks} where, we report the average Frobenius distance between true and estimated parameters for each mixture component. Notably, \textit{Sparsemixmat} outperforms the competing methods, exhibiting the lowest average distance for every mean matrix across all three clusters. While \textit{Sparsemixmat-lasso} and \textit{Full MGMM} seem to perform slightly better when looking at row and column precision matrices, the difference is often negligible. Moreover, our proposed approach achieves superior results in terms of recovery the underlying cluster partition, as measured by the adjusted Rand index \citep[ARI,][]{hubert:arabie:1985}, as well as overall model parsimony, quantified by the number of estimated parameters. \textit{Sparsemixmat} shows a higher ARI and a lower number of non-zero parameters compared to \textit{Full MGMM} and \textit{Sparsemixmat-lasso}. It is important to note that \textit{Full MGMM} does not employ any shrinkage, resulting in a total of $(K-1)+K(pq+p(p+1)/2+q(q+1)/2)$ estimated parameters in all cases.

\begin{table}[t]
\centering
\caption{{\em Alternated-blocks row precision matrices} scenario. Frobenius distance between true and estimated parameters, adjusted Rand index (ARI), and number of non-zero parameters ($d_0$) averaged over $100$
repetitions. Bold numbers indicate the best performing method according to the considered metric. Standard errors are reported in brackets.}
\label{tab:sim_blocks}
\begin{tabular}{l|lll}
  \hline
 & \textit{Full MGMM} & \textit{Sparsemixmat} & \textit{Sparsemixmat-lasso} \\ 
   \hline
  $||\M_1-\hat{\M}_1||_F$ & 38.617 (77.98) & \textbf{32.965 (76.716)} & 34.624 (77.839) \\ 
  $||\M_2-\hat{\M}_2||_F$ & 36.773 (78.055) & \textbf{13.876 (37.567)} & 14.678 (37.592) \\ 
  $||\M_3-\hat{\M}_3||_F$ & 16.382 (30.183) & \textbf{7.714 (17.721)} & 8.161 (18.563) \\ 
  $||\OMEGA_1-\hat{\OMEGA}_1||_F $& \textbf{1.136 (0.97)} & 3.218 (0.712) & 3.028 (0.647) \\ 
  $||\OMEGA_2-\hat{\OMEGA}_2||_F$ & \textbf{1.256 (1.573)} & 1.47 (0.412) & 1.383 (0.393) \\ 
  $||\OMEGA_3-\hat{\OMEGA}_3||_F$ & 1.529 (2.07) & 0.796 (0.513) & \textbf{0.75 (0.494)} \\ 
  $||\GAMMA_1-\hat{\GAMMA}_1||_F$ & \textbf{2.767 (6.117)} & 2.807 (6.002) & 2.797 (6.095) \\ 
  $||\GAMMA_2-\hat{\GAMMA}_2||_F$ & 3.794 (6.661) & 2.272 (4.251) & \textbf{2.239 (4.336)} \\ 
  $||\GAMMA_3-\hat{\GAMMA}_3||_F$ & 5.376 (10.043) & 4.24 (8.431) & \textbf{4.202 (8.609)} \\ 
  ARI & 0.948 (0.156) & \textbf{0.992 (0.058)} & 0.991 (0.058) \\ 
  $d_0$ & 362 (0) & \textbf{166.602 (19.931)} & 175.913 (15.51) \\ 
   \hline
\end{tabular}
\end{table}

\begin{figure}
\centering
    \includegraphics[scale=.8]{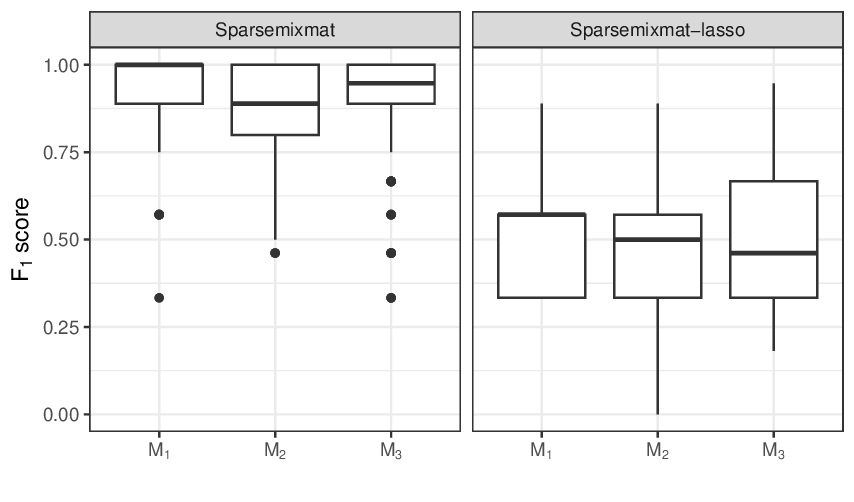}
    \caption{{\em Alternated-blocks row precision matrices} scenario. Boxplots of the $F_1$ score for $100$ replications of the experiment.}
  \label{fig:F1_M_block_cov}
\end{figure}

Another aspect to examine is the performance of the proposed approach in terms of variable selection. Specifically, given the matrix-variate nature of the data, we are interested in monitoring the method's ability in correctly identifying the zero rows of the mean matrices, and hence correctly detect those variables that are constantly equal to zero across occasions and clusters. To measure this, we make use of the $F_1$ score, defined as follows:
\begin{equation} \label{eq:f1_measure}
F_1=\frac{\texttt{tp}}{\texttt{tp}+0.5(\texttt{fp}+\texttt{fn})},
\end{equation}
where $\texttt{tp}$ denotes the number of zero rows in $\M_k$ correctly estimated as such, while$\texttt{fp}$ and $\texttt{fn}$ denote the number of non-zero rows wrongly shrunk to $0$ and the number of zero rows not shrunk to $0$, respectively. Figure \ref{fig:F1_M_block_cov} displays boxplots of the $F_1$ score for the \textit{Sparsemixmat} and \textit{Sparsemixmat-lasso} methods.
By enforcing entire rows of $\hat{\M}_k$ to be shrunk to zero by means of the group-lasso penalty, the {\em Sparsemixmat} approach achieves better variable selection performance. Conversely, for the {\em Sparsemixmat-lasso}, which applies entry-wise lasso shrinkage, there is no guarantee that entire rows will be ultimately set to $0$. Therefore, when the primary aim is multivariate variable selection or solving the support union problem within a matrix mixture context, our proposed approach is preferable. 

Similar results are observed when more complex dependence structures between the $p$ variables are considered, as it will be reported in the next subsection.

\begin{figure}
\centering
    \includegraphics[width=\linewidth]{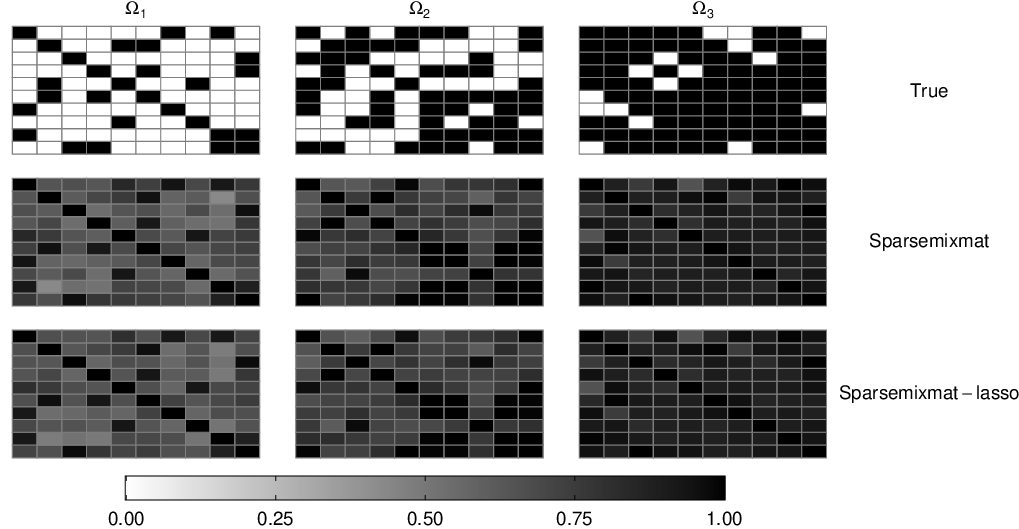}
    \caption{{\em Sparse-at-random row precision matrices} scenario. True association structures (top) and estimated association structures averaged over $100$ replications (middle and bottom) for the row precision matrices $\OMEGA_k$, $k=1,2,3$. Black squares denote a non-zero parameter between two variables.}
  \label{fig:Omega_graph_sparse_cov}
\end{figure}

\begin{table}[t]
\centering
\caption{{\em Sparse-at-random row precision matrices} scenario. Frobenius distance between true and estimated parameters, adjusted Rand index (ARI), and number of non-zero parameters ($d_0$) averaged over $100$ repetitions. Bold numbers indicate the best performing method according to the considered metric. Standard errors are reported in brackets.}
\label{tab:sim_sparse}
\begin{tabular}{l|lll}
  \hline
 & \textit{Full MGMM} & \textit{Sparsemixmat} & \textit{Sparsemixmat-lasso} \\ 
   \hline
  $||\M_1-\hat{\M}_1||_F$ & 60.188 (93.856) & \textbf{51.972 (90.38)} & 53.639 (91.107) \\ 
  $||\M_2-\hat{\M}_2||_F$ & 39.912 (58.11) & \textbf{17.181 (31.259)} & 20.183 (29.645) \\ 
  $||\M_3-\hat{\M}_3||_F$ & 37.871 (45.368) & \textbf{11.685 (21.207)} & 12.44 (22.506) \\ 
  $||\OMEGA_1-\hat{\OMEGA}_1||_F $& 4.704 (7.408) & 3.19 (5.278) & \textbf{3.152 (5.35)} \\ 
  $||\OMEGA_2-\hat{\OMEGA}_2||_F$ & 5.277 (7.607) & 4.353 (5.774) & \textbf{4.233 (5.758)} \\ 
  $||\OMEGA_3-\hat{\OMEGA}_3||_F$ & 6.525 (9.775) & 5.694 (5.622) & \textbf{5.628 (5.666)} \\ 
  $||\GAMMA_1-\hat{\GAMMA}_1||_F$ & 2.623 (4.448) & \textbf{2.884 (6.512)} & 2.981 (6.661) \\ 
  $||\GAMMA_2-\hat{\GAMMA}_2||_F$ & 12.287 (23.165) & \textbf{9.49 (18.434)} & 11.866 (22.579) \\ 
  $||\GAMMA_3-\hat{\GAMMA}_3||_F$ & 18.079 (26.515) & 17.61 (23.537) & \textbf{16.404 (24.64)} \\ 
  ARI & 0.944 (0.162) & \textbf{1 (<0.01)} & \textbf{1 (<0.01)} \\ 
  $d_0$ & 362 (0) & 257.204 (20.289) & \textbf{251.071 (12.01)} \\ 
   \hline
\end{tabular}
\end{table}

\begin{figure}
\centering
    \includegraphics[scale=.8]{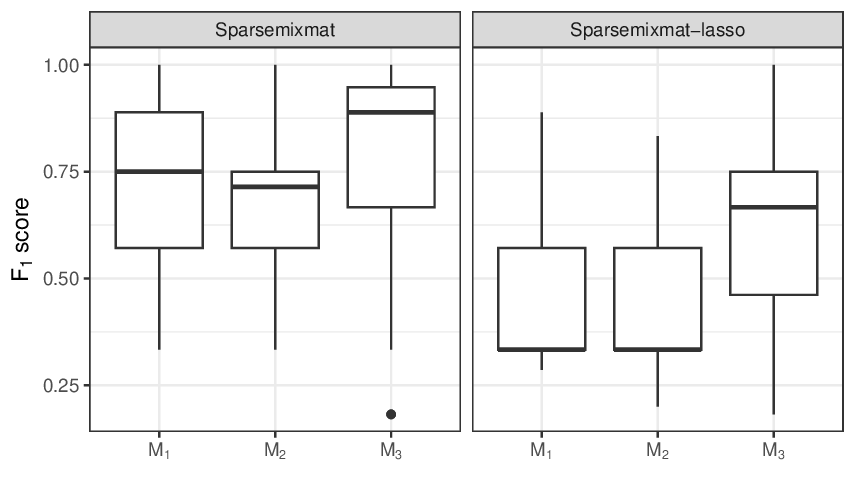}
    \caption{{\em Sparse-at-random row precision matrices} scenario. Boxplots of the $F_1$ score for $100$ replications of the experiment.}
  \label{fig:F1_M_sparse_cov}
\end{figure}

\subsubsection{Sparse-at-random row precision matrices}
In the second scenario, the row precision matrices are constructed having a sparse-at-random Erd\H{o}s-R\'{e}nyi graph structure. Figure \ref{fig:Omega_graph_sparse_cov} shows the data-generating and estimated association structures, with interpretation similar to previous similar figures. We note how the more challenging dependence patterns among the variables affect the performance of the penalized models. Irrespective of the considered methods, the number of non-zero parameters is consistently overestimated, resulting in solutions where the levels of sparsity of the $\OMEGA_k$ matrices are underestimated. Similar results are observed for the column precision matrices and the mean matrices (not reported here). Nonetheless, {\em Sparsemixmat} seems to outperform {\em Full MGMM} and {\em Sparsemixmat-lasso} when evaluating the performance in terms of Frobenius distance and recovering of the true clustering, as it is indicated in Table \ref{tab:sim_sparse}). Particularly, the mean and column-precision matrices are quite satisfactorily estimated by {\em Sparsemixmat}, with only a slightly higher total number of parameters in comparison to \textit{Sparsemixmat-lasso}. Similarly to the previous scenario, with regard to the ability of performing variable selection, a group-lasso penalty on the rows of $\M_k$ is to be preferred, as highlighted in the boxplots of Figure \ref{fig:F1_M_sparse_cov}, where {\em Sparsemixmat} shows consistently higher $F_1$ score values compared to {\em Sparsemixmat-lasso}. Interestingly, the variable selection performance of both methods in terms of the $F_1$ score is lower in this scenario compared to the previous one. This finding suggests that the performance in variable selection does not only depend on the penalty imposed to the mean matrices, but it is also affected by how well the dependence structure among the $p$ variables in the $K$ clusters is recovered.

In summary, the proposed approach adequately tackle the problem of clustering matrix-variate data with sparse model parameters. The method is flexible, it is capable of capturing cluster-wise different dependence structures in both variables and occasions, it enables row-wise variable selection when variables are recorded over multiple occasions, and it detects effectively the clustering structure in the matrix data. These considerations hold true not only in an experimental setup but also in the analysis of real-world data, as reported in the next section.
 





\section{Application: criminal trends in the US}\label{sec:crimeApplication}
\subsection{Data description}
We analyze data from the United States Department of Justice Federal Bureau of Investigation concerning violent and property crimes of $236$ American cities. The aim of the analysis is to cluster cities with similar crime trends and to identify which crime types exhibit relevant differences in the time patterns across clusters. In the data, for each city, $p=7$ variables capturing the rates of murder, rape, robbery, aggravated assault, burglary, larceny-theft, and motor vehicle theft are measured over $q=13$ years in the period between $2000$ and $2012$. Thus, the data can be conveniently arranged in a $7 \times 12 \times 236$ array, where each statistical unit $\X_i$, $i=1,\ldots,236$, takes the form of a $7 \times 13$ matrix. The dataset is publicly available within the \texttt{MatTransMix R} package \citep{Zhu2022} and has been previously analyzed in \cite{Melnykov2019a}, where the authors introduced a method based on mixture of matrix transformation regression time series.  The next subsection includes discussion of the results of our modeling approach and comparisons with the findings from \cite{Melnykov2019a}.

\subsection{Results}
We implement an initial pre-processing step in which the statistical units are cell-wise centered and log-transformed to alleviate skewness. Subsequently, the {\em Sparsemixmat} model introduced in Section \ref{sec:sparsemixmat} is fitted to the crime data. The shrinkage parameters are varied within a pre-specified grid of values, and considering $K \in\{3,4,5,6\}$. 

\begin{figure}
\centering
    \includegraphics[width=\linewidth]{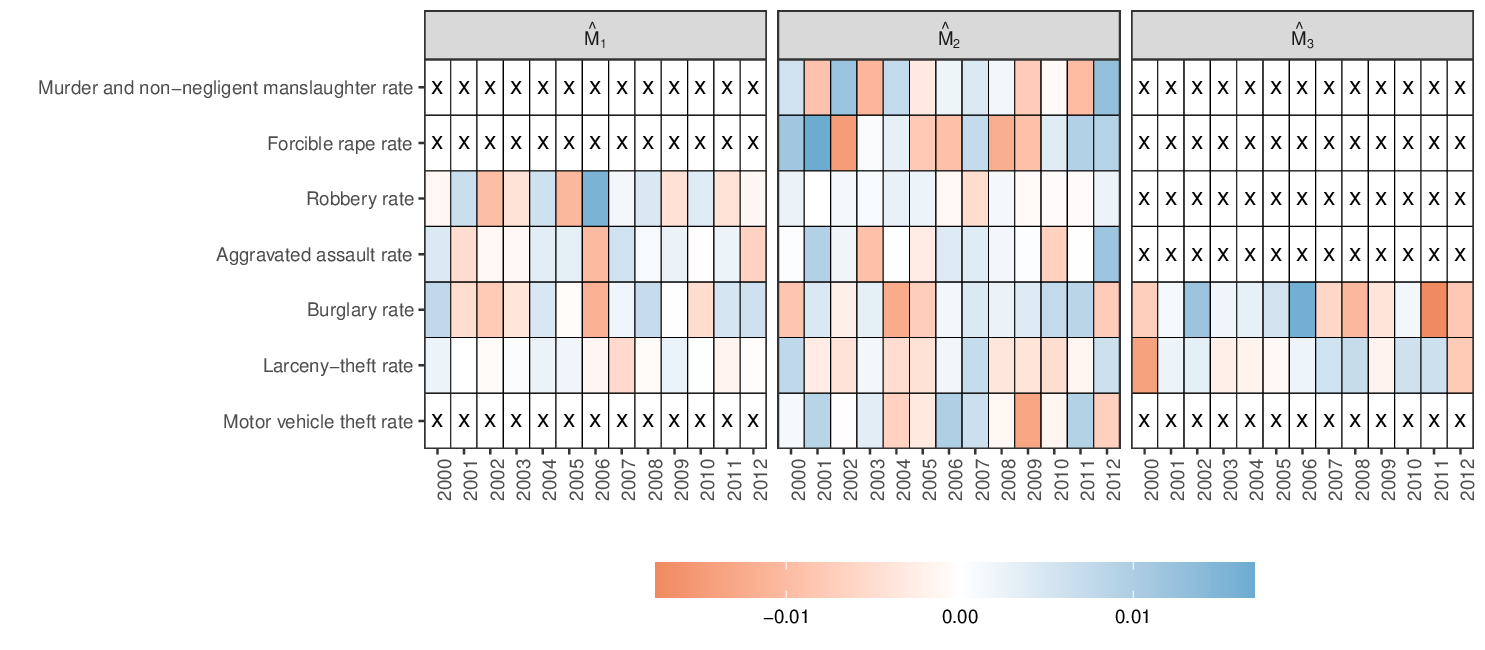}
    \caption{{\em Crime data}. Estimated mean matrices $\hat{\M}_k$, $k=1,2,3$ for the \textit{sparsemixmat} model. Colors denote the values of the estimates; a $0$ entry in the matrices is indicated by the symbol $\times$.}
  \label{fig:gr_lasso_M_crime}
\end{figure}

\begin{figure}
\centering
    \includegraphics[width=\linewidth]{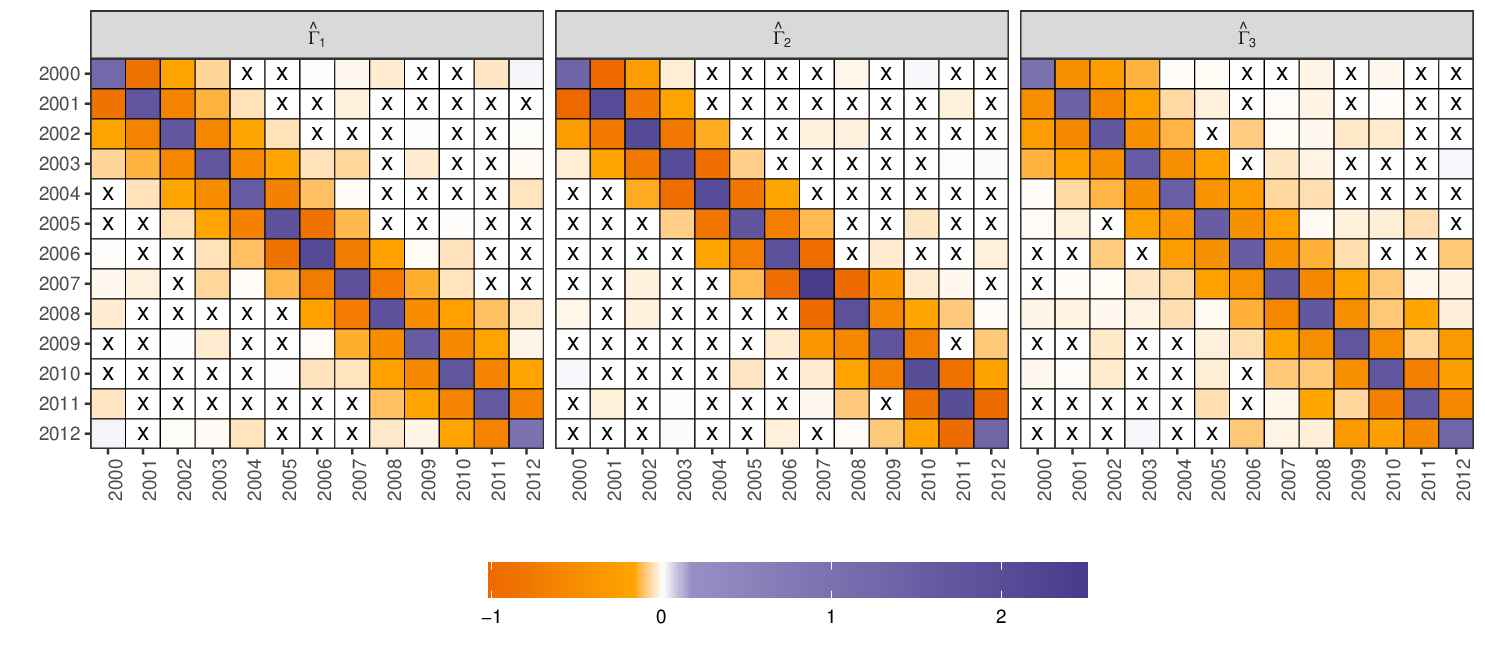}
\caption{{\em Crime data}. Estimated column precision matrices $\hat{\GAMMA}_k$, $k=1,2,3$, for the \textit{sparsemixmat} model. Colors denote the values of the entries; a $0$ entry in the matrices is indicated by the symbol $\times$.}
  \label{fig:gr_lasso_GAMMA_crime}
\end{figure}

The BIC as introduced in Equation \eqref{eq:modBIC} selects $K=3$ clusters, with corresponding shrinkage hyperparameters $\lambda_1$, $\lambda_2$, and $\lambda_3$ equal to $3.81$, $0$ and $14.3$, respectively. The penalty coefficient $\lambda_2 = 0$ implies that the estimated row precision matrices $\hat{\OMEGA_k}$ for the selected model are non-sparse, indicating  relevant associations between the crime types across clusters. On the other hand, with the selected $\lambda_1$ and $\lambda_3$ greater than zero, both the estimated mean matrices and the column precision matrices measuring the dependence between the time occasions have certain degrees of sparsity. Visual representations of these estimated parameters are displayed in figures \ref{fig:gr_lasso_M_crime} and \ref{fig:gr_lasso_GAMMA_crime}. From Figure \ref{fig:gr_lasso_M_crime}, we observe that no crime type presents estimated cluster mean rates equal to zero across all clusters, indicating that all variables contain some discriminating information. However, in light of the considerations of Section~\ref{sec:grLasso_modSpec}, clusters tend to be differentiated over the rates of certain crimes across the years. For example, all clusters have dissimilar burglary and larceny-theft rates, while cluster $1$ and $3$ tend to overlap in terms of murder, rape, and motor vehicle theft rates. In addition, robbery and assault crime rates tend to stay constant over time for the cities in cluster 3, while they vary for those in clusters 1 and 2. Figure \ref{fig:gr_lasso_GAMMA_crime} shows that the estimated column precision matrices, which embed the conditional association structure of the crime rates between years, tend to have a banded structure. The entries along the diagonal are generally non-zero, while entries between far in time occasions are generally shrunk to zero, indicating higher levels of association between consecutive years. 

\begin{figure}
\centering
    \includegraphics[width=\linewidth]{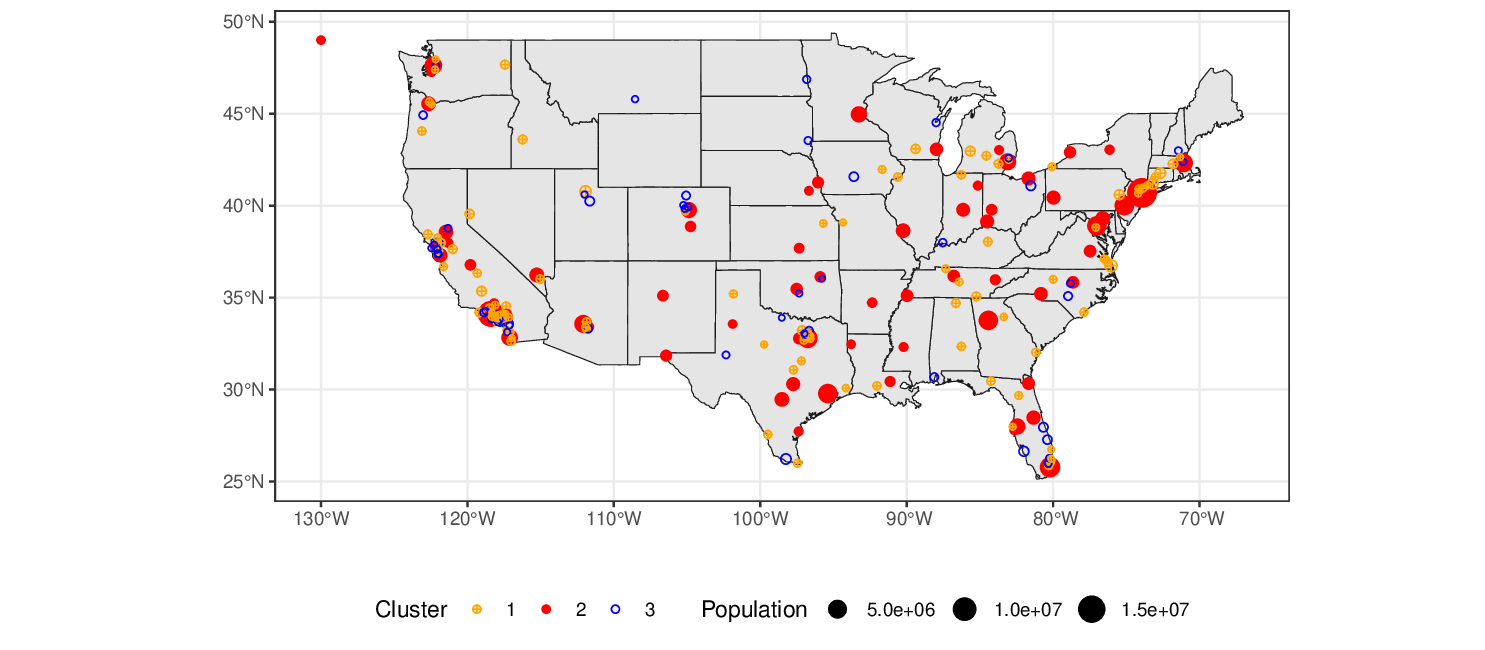}
\caption{Crime data. Map of the USA showing the clustering of the cities obtained from the \textit{sparsemixmat} model. The sizes of the circles is proportional to the city population. Colors and symbols indicate different clusters.}
  \label{fig:gr_lasso_map_crime}
\end{figure}

\begin{figure}
\centering
    \includegraphics[scale=.9]{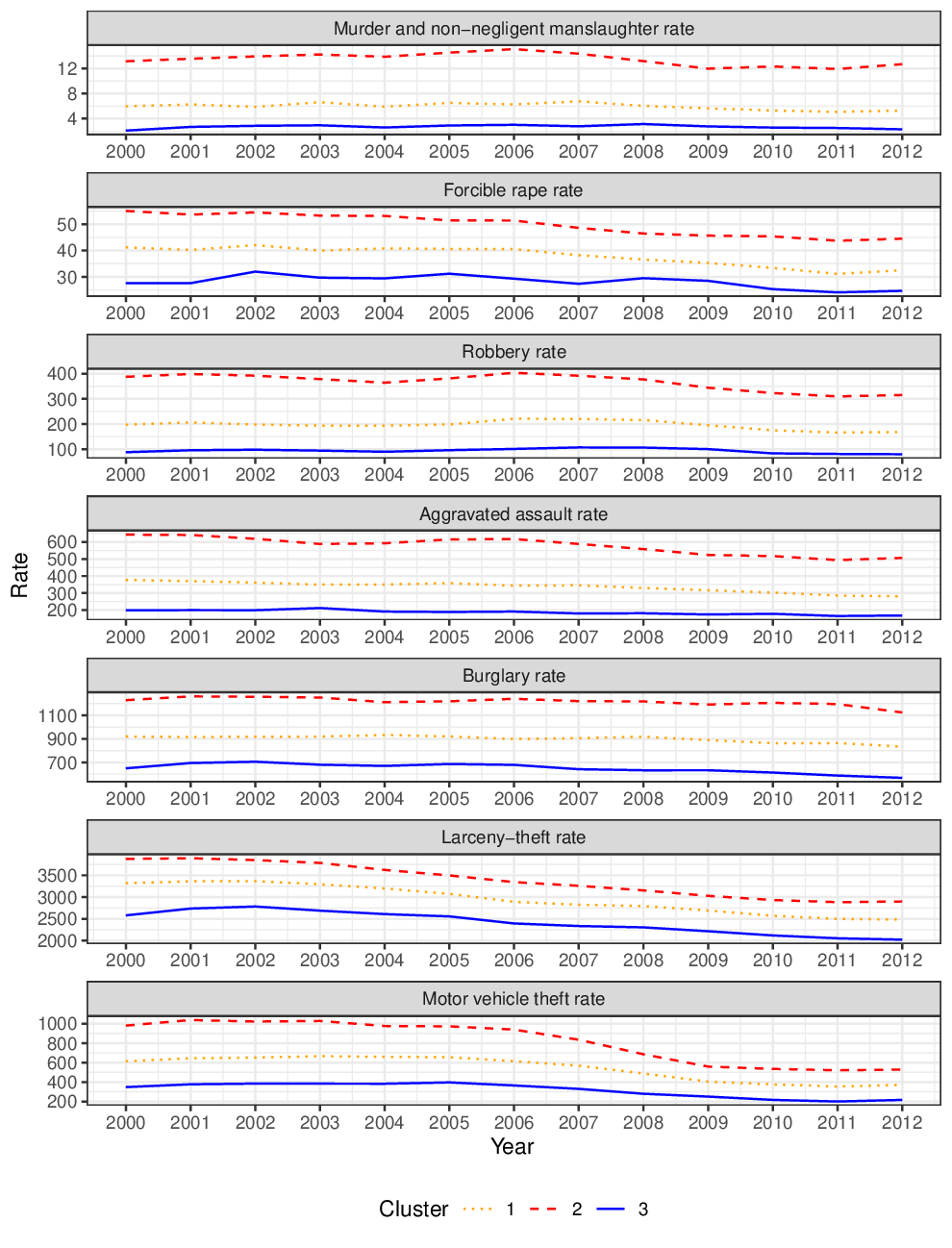}
\caption{Crime data. Mean profiles for the \textit{sparsemixmat} model. The mean profiles are computed for the variables in original scale. Colors and line types illustrate different clusters.}
  \label{fig:gr_lasso_mean_profiles_crime}
\end{figure}

The clustering of the cities in the data is displayed in the map of Figure \ref{fig:gr_lasso_map_crime}, while the mean rate profiles of the resulting partition, computed for the crime types in the original scale, are reported in Figure \ref{fig:gr_lasso_mean_profiles_crime}. More in detail, cluster 3 (blue color) identifies the safest cities in the country, which tend also to be the smallest in size. Higher concentration of safe cities can be observed in Northern Texas, the Los Angeles-San Diego area, and parts of the northern states, together with a few coastal areas in Florida and in the south of Indiana. Cluster 2 (red color) includes the cities with the highest crime rates of the considered types. From the map, it appears that these cities tend to be unevenly distributed across the US, with a concentration in the eastern part of the country, which is also the most densely populated. Lastly, cluster 1 (orange color) comprises cities that are slightly less safe, for which the mean crime rates over time tend to be higher than those in cluster 3. However, as remarked previously, the cities in these clusters tend to overlap in terms of murder, rape, and motor vehicle theft rates over time (see Figure \ref{fig:gr_lasso_M_crime}).

We highlight several similarities in the results discussed here and those of the analyses reported in \cite{Melnykov2019a}. First off, compared to the partition obtained in their $3$-cluster model we observe an agreement of approximately  $75 \%$ of cases, along with a very similar interpretation of the resulting clusters. 
Dependence patters across time similar to those displayed in Figure \ref{fig:gr_lasso_GAMMA_crime} have also been observed in \cite{Melnykov2019a}, in which a first order autoregressive model was employed to reduce the number of parameters and model the time dependence. While this is indeed a sensible modeling choice given the temporal dependence of these data, we remark the flexibility of our procedure in automatically capturing an autoregressive-like structure in the time occasions. This is achieved through the penalization imposed on the column precision matrices without the need to pre-specify any pattern or dependence structure. In addition, in \cite{Melnykov2019a}, to overcome the overparametrization issue associated to the mean matrices, the authors consider regressing crime rates on years. In contrast, our proposed method employs a group-lasso penalty which effectively serves the same purpose, without the specification of a regression model.
\begin{figure}
\centering
    \includegraphics[width=\linewidth]{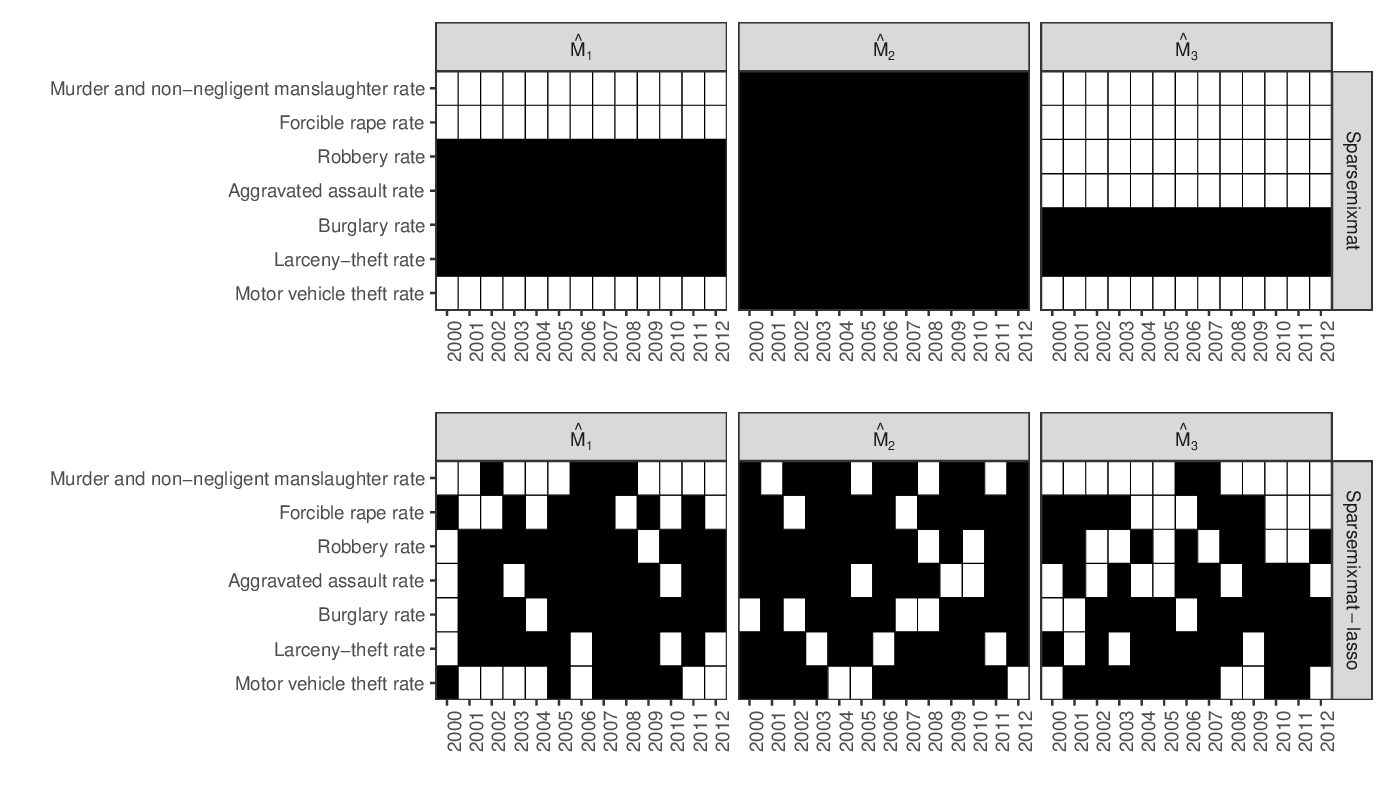}
    \caption{{\em Crime data}. Sparse structure associated to the mean matrices $\M_k$, $k=1,2,3$ for the \textit{sparsemixmat} and \textit{sparsemixmat-lasso} models. Black squares denote an entry different from $0$.}
  \label{fig:M_crime_gr_lasso_lasso.eps}
\end{figure}

We conclude this section by comparing the results obtained with our \textit{sparsemixmat} procedure with the \textit{sparsemixmat-lasso} of \cite{heo2021penalized}. The two models provide very similar partitions of the cities in the data, having almost perfect agreement and with only $9$ cities assigned to different clusters. Figure \ref{fig:M_crime_gr_lasso_lasso.eps} shows the estimated sparse structures associated with the estimated cluster mean matrices, obtained under the two different penalties. For {\em Sparsemixmat-lasso}, all the crime types have non-zero mean rates for some of the years and clusters, making difficult to differentiate the clusters in terms of overall mean crime rate patters across years. Once again, it is worth to highlight the ease of interpretation induced by the group-lasso penalty of {\em Sparsemixmat}, making it a more favorable option when clustering with matrix-variate data where variables are recorded over multiple time occasions.



\section{Conclusion} \label{sec:conc}

The complex structure entailed by three-way data makes clustering matrices a particularly challenging task. By framing the problem into a well-defined probabilistic context, model-based approaches are unarguably among the most commonly adopted to address these challenges. Nonetheless, these approaches have to face severe issues and limitations even when dealing with three-way data of moderate dimensions. In this work, we propose a modeling framework that alleviates these drawbacks, thus allowing to cluster matrix-variate data even when the number of variables $p$ or the number of occasions $q$ is moderate. In particular, the presented method relies on a penalized likelihood approach that allows to induce sparsity in the model parameters. The penalties on the row and column precision matrices reduce greatly the number of parameters to estimate, while simultaneously easing the interpretation of the dependence patterns, thanks to the connection with Gaussian graphical models. Additionally, the group lasso penalty on the rows of the component mean matrices allows to perform variable selection in the situation where the three-way data arise from variables recorded over multiple occasions. This increases even further the model parsimony and provides useful indications regarding the ability of the variables in separating the clusters across the occasions. Assessments on both synthetic data and data concerning crime rates in the US have shown the validity of our proposed method for sparse model-based clustering of three-way data, overcoming some of the drawbacks of the approaches currently present in the literature. 

The paper leaves several paths open for future research. Firstly, while effectively performing variable selection, the group lasso penalty, could be quite rigid in some applications. In fact, as highlighted in Section \ref{sec:grLasso_modSpec}, this specification sets to zero entire rows of the mixture component mean matrices. Nonetheless, sometimes sparsity could be desirable also within the rows, thus enforcing only some elements and not the entire variable to be shrunk to zero. 
This could be achieved by adapting the so-called sparse group lasso \citep{simon2013sparse} to the framework considered in our work. In fact, this penalty is a convex combination of the group-lasso and the entry-wise lasso penalty briefly described in Section \ref{sec:Lasso_modSpec}, and it could extend the application of the proposed framework to other contexts. Throughout the manuscript, matrix Gaussian mixture models have been parameterized in terms of precision matrices. Nonetheless, the penalized approach can be adapted to a setting where sparsity is imposed on the covariance matrices, thus generalizing the work by \citet{fop:2019} to the matrix-variate case. This approach would still lend itself to a convenient representation in terms of the so-called \emph{covariance graphs}, where a missing edge between two nodes implies that the corresponding variables are marginally independent \citep{chaudhuri:etal:2007}. 
Furthermore, in this work we focused on matrix Gaussian distributions, since they are a widely adopted choice to model continuous data. Nonetheless, it would be interesting to explore if the proposed penalized method could be employed in conjunction with other choices for the component densities, potentially encompassing situations with heavy-tails or skewness \citep[see e.g.,][]{melnykov2018model, tomarchio2020two}. Lastly, alternative model selection strategies might be devised. The adopted grid search produced good results in our numerical assessments. Nonetheless, as mentioned in Section \ref{sec:model_selection}, it could be too computationally demanding in some applications. For this reason, stochastic optimization techniques could be borrowed and adapted to our setting, as well as the so-called E-MS algorithm introduced by \citet{jiang2015ms}. 

As a final worthy observation, we noted that even in the matrix-variate scenario, the works focusing on precision matrices estimation in multi-class settings often enforce similarities between the underlying graphical models \citep{Huang2015}. This assumption, reasonable in different applications,  could deteriorate the quality of the results when clustering is the final aim. Therefore, we believe that the strategy adopted in \citet{casa2022group} could be combined with the procedure proposed in this paper, to encompass those situations where different component precision matrices have markedly different degrees of sparsity. 

\section*{Acknowledgments}
Andrea Cappozzo acknowledges the support by MUR, grant Dipartimento di Eccellenza 2023-2027. 

\section*{Conflicts of interest}
The authors report there are no competing interests to declare.

\begin{center}
{\large\bf SUPPLEMENTARY MATERIAL}
\end{center}
The supplementary material reports the proof of Proposition 1.
\begin{proof}[Proof of Proposition 1]
For easying the notation, we subsequently drop the ``hat'' from any parameter estimate and, without loss of generality, we prove the result for $\PP_1$ equal to an all-ones matrix. 
Similarly to the case outlined in Theorem 1 of \cite{zhou:2009}, $Q_M(\M_k)$ is differentiable with respect to $m_{ls,k}$ when $m_{ls,k} \neq 0$, while non-differentiable at $m_{ls,k}=0$. The following two cases are considered:\\
\begin{enumerate}
\item If $m_{ls,k} \neq 0$ is a maximum, given that $Q_{M}(\M_k)$ is concave and differentiable, the sufficient and necessary condition for $m_{ls,k}$ to be the global maximum of $Q_{M}(\M_k)$ is
\begin{equation} \label{eq:deriv_m_gr_0}
 \frac{\partial Q_{M}(\M_k)}{\partial m_{ls,k}}=0 \Longleftrightarrow  \sum_{i=1}^n z_{ik}\sum_{r=1}^p\sum_{c=1}^q\omega_{lr,k} x_{rc,i}\gamma_{cs,k}-n_k\sum_{r=1}^p \sum_{c=1}^q\omega_{lr,k}m_{rc,k}\gamma_{cs,k}-\lambda_1\sign(m_{ls,k})=0
\end{equation}
from which \eqref{eq:m_update_m_gr_0} is easily derived by solving \eqref{eq:deriv_m_gr_0} with respect to $m_{ls,k}$.
\item If $m_{ls,k} = 0$ is a maximum, we compare $Q_{M}(0,\cdot)$ with $Q_{M}(\Delta m_{ls,k},\cdot)$, the values of  $Q_{M}(\M_k)$ at $m_{ls,k}=0$ and $m_{ls,k}=\Delta m_{ls,k}$ respectively (while the other entries of $\M_{k}$ are fixed at their maximum). By definition, we have
$Q_{M}(0, \cdot) \geq Q_{M}\left(\Delta m_{ls,k}, \cdot\right)$ for any $\Delta m_{ls,k}$ near $0$\\

$\Longleftrightarrow$
\begin{multline*}
\sum_{i=1}^n z_{ik} \left[ -2 \left. \tr \left\{ \OMEGA_k\X_i\GAMMA_k\M_k^{'}\right\} \right |_{m_{ls,k}=\Delta m_{ls,k}}+ \left.\tr \left\{ \OMEGA_k\M_k\GAMMA_k\M_k^{'}\right\} \right |_{m_{ls,k}=\Delta m_{ls,k}} \right. +\\ \left.+ 2 \left. \tr \left\{ \OMEGA_k\X_i\GAMMA_k\M_k^{'}\right\} \right |_{m_{ls,k}=0} - \left.\tr \left\{ \OMEGA_k\M_k\GAMMA_k\M_k^{'}\right\} \right |_{m_{ls,k}=0}\right] \geq - 2 \lambda_1 |\Delta m_{ls,k}|
\end{multline*}
$\Longleftrightarrow$
\begin{multline*}
\sum_{i=1}^n z_{ik} \left[ -2 \left( \left. \tr \left\{ \OMEGA_k\X_i\GAMMA_k\M_k^{'} \right |_{m_{ls,k}=\Delta m_{ls,k}} -\left. \OMEGA_k\X_i\GAMMA_k\M_k^{'}\right |_{m_{ls,k}=0}\right\} \right) \right. +\\
+ \left. \left. \tr \left\{ \OMEGA_k\M_k\GAMMA_k\M_k^{'} \right |_{m_{ls,k}=\Delta m_{ls,k}} -  \left. \OMEGA_k\M_k\GAMMA_k\M_k^{'}\right |_{m_{ls,k}=0} \right\} \right] \geq - 2 \lambda_1 |\Delta m_{ls,k}|
\end{multline*}
$\Longleftrightarrow$
\begin{multline*}
\sum_{i=1}^n z_{ik} \left[ -2 \left[ \sum_{r=1}^p \omega_{rl,k} \left( \sum_{c=1}^qx_{rc,i}\gamma_{cs,k}\right)\Delta m_{ls,k}\right] \right. +\\
 \left. 2 \sum_{\substack{r=1 \\ r\neq l}}^p \omega_{rl,k} \left( \sum_{c=1}^qm_{rc,k}\gamma_{cs,k}\right)\Delta m_{ls,k}+\omega_{ll,k}\left( \Delta m_{ls,k} \gamma_{ss,k}+2\sum_{\substack{c=1 \\ c\neq s}}^q m_{lc,k}\gamma_{cs,k}\right)\Delta m_{ls,k}  \right] \geq - 2 \lambda_1 |\Delta m_{ls,k}|
\end{multline*}
$\Longleftrightarrow$
\begin{multline*}
\sum_{i=1}^n z_{ik} \left[ 2 \left[ \sum_{\substack{r=1 \\ r\neq l}}^p \omega_{rl,k} \left( \sum_{c=1}^qx_{rc,i}\gamma_{cs,k}\right)\Delta m_{ls,k}\right] +2 \omega_{ll,k} \left( \sum_{c=1}^qx_{lc,i}\gamma_{cs,k}\right) \Delta m_{ls,k}\right. -\\
 \left. 2 \sum_{\substack{r=1 \\ r\neq l}}^p \omega_{rl,k} \left( \sum_{c=1}^qm_{rc,k}\gamma_{cs,k}\right)\Delta m_{ls,k}-\omega_{ll,k}\left( \Delta m_{ls,k} \gamma_{ss,k}+2\sum_{\substack{c=1 \\ c\neq s}}^q m_{lc,k}\gamma_{cs,k}\right)\Delta m_{ls,k}  \right] \leq  2 \lambda_1 |\Delta m_{ls,k}|
\end{multline*}
\end{enumerate}
$\Longleftrightarrow$
\begin{multline*}
\sum_{i=1}^n z_{ik} \left[ 2 \Delta m_{ls,k} \left[ \sum_{\substack{r=1 \\ r\neq l}}^p \omega_{rl,k} \left( \sum_{c=1}^q\left(x_{rc,i}-m_{rc,k}\right)\gamma_{cs,k}\right)\right] +2 \omega_{ll,k} \left( \sum_{c=1}^qx_{lc,i}\gamma_{cs,k}\right) \Delta m_{ls,k}\right. -\\
\left. \omega_{ll,k}\left( \Delta m_{ls,k} \gamma_{ss,k}+2\sum_{\substack{c=1 \\ c\neq s}}^q m_{lc,k}\gamma_{cs,k}\right)\Delta m_{ls,k}  \right] \leq  2 \lambda_1 |\Delta m_{ls,k}|
\end{multline*}
$\Longleftrightarrow$
\begin{multline*}
\left| \sum_{i=1}^n z_{ik} \left[  \sum_{\substack{r=1 \\ r\neq l}}^p \omega_{rl,k} \left( \sum_{c=1}^q\left(x_{rc,i}-m_{rc,k}\right)\gamma_{cs,k}\right) + \omega_{ll,k} \left( \sum_{c=1}^qx_{lc,i}\gamma_{cs,k}\right) \right. \right. \\- \left. \left.  \omega_{ll,k}\left( \frac{\Delta m_{ls,k}}{2} \gamma_{ss,k}+\sum_{\substack{c=1 \\ c\neq s}}^q m_{lc,k}\gamma_{cs,k}\right) \right] \right| \leq   \lambda_1
\end{multline*}
$\Longleftrightarrow$
\begin{multline*}
\left| \sum_{i=1}^n z_{ik} \left[  \sum_{\substack{r=1 \\ r\neq l}}^p \omega_{rl,k} \left( \sum_{c=1}^q\left(x_{rc,i}-m_{rc,k}\right)\gamma_{cs,k}\right) + \right.\right.\\
  \left. \omega_{ll,k} \left( \sum_{\substack{c=1 \\ c\neq s}}^q\left(x_{lc,i}-m_{lc,k}\right)\gamma_{cs,k}\right) \right. +  \left. \omega_{ll,k}x_{ls,i}\gamma_{cs,k} \Bigg] \right| \leq   \lambda_1 \text{ as }  \Delta m_{ls,k}\rightarrow 0
\end{multline*}

\end{proof}
%
%
%
%
%

\bibliographystyle{apalike}
\bibliography{bibliography}

\end{document}